\documentclass{aa}
\usepackage{graphicx}
\usepackage[varg]{txfonts}

\usepackage{natbib}
\bibpunct{(}{)}{;}{a}{}{,}

\usepackage{color}
\usepackage{mathrsfs}
\usepackage{mathtools}
\definecolor{comm}{rgb}{0,0.5,0}

\definecolor{old}{rgb}{0.6,0.4,0.4}

\definecolor{new}{rgb}{0.5,0.2,0.5}

\definecolor{done}{rgb}{1,0.5,0}

\definecolor{todo}{rgb}{0.8,0,0}

\definecolor{idea}{rgb}{0.3,0.5,1}

\newcommand{\cd}{d$^{-1}$}
\allowdisplaybreaks

\begin{document} 

\title{Theoretical investigation of the occurrence of tidally excited oscillations in massive eccentric binary systems}
  \author{P. A. Ko{\l}aczek-Szyma\'nski
  \and
  T. R\'{o}\.{z}a\'nski
}
  \institute{Astronomical Institute, University of Wroc\l aw, Kopernika 11, 51-622 Wroc\l aw, Poland\\
  \email{kolaczek@astro.uni.wroc.pl}
}
  \date{Received 16.10.2022; Revised 01.12.2022; Re-revised 28.12.2022; Accepted 02.01.2023}
% \abstract{}{}{}{}{} 
% 5 {} token are mandatory
\abstract
{Massive and intermediate-mass stars reside in binary systems much more frequently than low-mass stars. At the same time, binaries containing massive main-sequence (MS) component(s) are often characterised by eccentric orbits, and can therefore be observed as eccentric ellipsoidal variables (EEVs). The orbital phase-dependent tidal potential acting on the components of EEV can induce tidally excited oscillations (TEOs), which can affect the evolution of the binary system.}
{We investigate how the history of resonances between the eigenmode spectra of the EEV components and the tidal forcing frequencies depends on the initial parameters of the system, limiting our study to MS. Each resonance is a potential source of TEO. We are particularly interested in the total number of resonances, their average rate of occurrence and their distribution in time.} 
{We synthesised 20,000 evolutionary models of the EEVs across the MS using Modules for Experiments in Stellar Astrophysics (\texttt{MESA}) software for stellar structure and evolution. We considered a range of masses of the primary component from 5 to 30\,$M_\odot$. Later, using the \texttt{GYRE} stellar non-adiabatic oscillations code, we calculated the eigenfrequencies for each model recorded by \texttt{MESA}. We focused only on the $l=2$, $m=0,+2$ modes, which are suspected of being dominant TEOs. Knowing the temporal changes in the orbital parameters of simulated EEVs and the changes of the eigenfrequency spectra for both components, we were able to determine so-called `resonance curves', which describe the overall chance of a resonance occurring and therefore of a TEO occurring. We analysed the resonance curves by constructing basic statistics for them and analysing their morphology using machine learning methods, including the Uniform Manifold Approximation and Projection (UMAP) tool.}
{The EEV resonance curves from our sample are characterised by striking diversity, including the occurrence of exceptionally long resonances or the absence of resonances for long evolutionary times. We found that the total number of resonances encountered by components in the MS phase ranges from $\sim$$10^2$ to $\sim$$10^3$, mostly depending on the initial eccentricity. We also noticed that the average rate of resonances is about an order of magnitude higher ($\sim$$10^2$\,Myr$^{-1}$) for the most massive components in the assumed range than for EEVs with intermediate-mass stars ($\sim$$10^1$\,Myr$^{-1}$). The distribution of resonances over time is strongly inhomogeneous and its shape depends mainly on whether the system is able to circularise its orbit before the primary component reaches the terminal-age MS (TAMS). Both components may be subject to increased resonance rates as they approach the TAMS. Thanks to the low-dimensional UMAP embeddings performed for the resonance curves, we argue that their morphology changes smoothly across the resulting manifold for different initial EEV conditions. The structure of the embeddings allowed us to explore the whole space of resonance curves in terms of their morphology and to isolate some extreme cases.}
% conclusions heading (optional)
{Resonances between tidal forcing frequencies and stellar eigenfrequencies cannot be considered rare events for EEVs with massive and intermediate-mass MS stars. On average, we should observe TEOs more frequently in EEVs containing massive components than intermediate-mass ones. TEOs will be particularly well-pronounced for EEVs with the component(s) close to the TAMS, which begs for observational verification. Given the total number of resonances and their rates, TEOs may play an important role in the transport of angular momentum within massive and intermediate-mass stars (mainly near TAMS).}
\titlerunning{Tidally excited oscillations in massive and intermediate-mass EEVs}
\authorrunning{Ko{\l}aczek-Szyma\'nski \& R\'{o}\.{z}a\'nski}
\maketitle
%
%-------------------------------------------------------------------

\section{Introduction}\label{sect:introduction}
For many reasons, massive stars ($\gtrsim 8\,M_\odot$) are of particular interest to modern astrophysics. Primarily, they are progenitors of core-collapse supernovae \citep[e.g.,][]{2007PhR...442...38J,2009ARA&A..47...63S} and long $\gamma$-ray bursts \citep[e.g.,][]{2006Natur.441..463F,2006A&A...460..199Y}. For billions of years, they contributed to the chemical evolution of the entire Universe and interacted mechanically with the surrounding interstellar medium \citep[e.g.,][]{2007ApJ...662.1268O,2012ApJ...759..108S}, also through their intense line-driven stellar winds \citep{2021arXiv210908164V}. Furthermore, most of their remnants are compact objects, such as neutron stars (NSs, and among them magnetars and pulsars) and black holes (BHs), which allow the empirical study of effects of general relativity. Finally, massive stars can be observed at cosmological distances due to their enormous luminosities, hence they dominate in the spectra of distant starburst galaxies \citep[see][for a recent review]{2022arXiv220201413E}. These features of massive stars (and many others) demonstrate that understanding the structure and evolution of massive stars is one of the key tasks of astronomy.

As is well known, massive and intermediate-mass ($\gtrsim 2\,M_\odot$ and $\lesssim 8\,M_\odot$) stars reside in binary systems much more frequently than their lower-mass counterparts \citep{2013ARA&A..51..269D,2012Sci...337..444S}. Moreover, as shown, for example, by \cite{2014ApJS..215...15S} and \cite{2017ApJS..230...15M}, O-type dwarfs in particular are often found in multiple systems. This shows that binarity is inherent in the evolution of massive stars and cannot be ignored when studying these objects as well as their final outcomes. Many interesting phenomena in the Universe are the result of binarity among massive and/or intermediate-mass stars. These include Be stars \citep{1975BAICz..26...65K,2020A&A...641A..42B}, so-called `stripped stars' \citep{2020ApJ...904...56G,2020A&A...639L...6S,2022MNRAS.511L..24E}, BH-BH/BH-NS/NS-NS mergers \citep[progenitors of gravitational-wave events,][]{2016PhRvL.116f1102A,2019PhRvX...9c1040A}, `early' stellar mergers \citep{2020MNRAS.491.5158T,2022A&A...659A..98S,2022ApJ...932...14L}, Ib/c supernova progenitors \citep{2012ARA&A..50..107L,2012A&A...544L..11Y}, `massive Algols' \citep{2007A&A...467.1181D,2017AcA....67..329S,2022A&A...659A..98S}, and even Wolf-Rayet stars \citep{2019A&A...627A.151S,2022A&A...659A...9P}.

At the same time, binary systems that contain massive main-sequence (MS) component(s) are often characterised by eccentric orbits due to their relatively young age and the presence of radiative outer layers, which are less vulnerable to tidal dissipation compared to convective envelopes \citep[e.g.,][]{2016ApJ...824...15V}. Both observational studies of large samples of massive binaries \citep[e.g.,][]{2017ApJS..230...15M} and hydrodynamical simulations of their formation \citep[see][and references therein]{2020A&A...644A..41O} suggest significantly non-zero eccentricities at their birth.

Assuming that the periastron distance between the components is sufficiently small\footnote{That is, of the order of a few radii of the larger component.}, the combined proximity effects, such as ellipsoidal distortion, irradiation/reflection effect and Doppler beaming/boosting, make such a system an eccentric ellipsoidal variable \citep[hereafter EEV, e.g.,][]{2012MNRAS.421.2616N}. Due to the characteristic shape of the light curve of EEV during the periastron passage (which can resemble an electrocardiogram), EEVs are sometimes referred to as `heartbeat stars' \citep{2011ApJS..197....4W,2012ApJ...753...86T,2014A&A...564A..36B,2016AJ....151...68K,2021A&A...647A..12K,2022ApJS..259...16W}. 

The orbital phase-dependent tidal potential acting on the components of EEV can induce tidally excited oscillations (TEOs) in their interiors \citep{1975A&A....41..329Z,1995ApJ...449..294K,2017MNRAS.472.1538F,2021FrASS...8...67G}, which in turn can affect the evolution of the binary system. However, many details of TEOs in massive and intermediate-mass stars are still poorly understood including the total number of TEOs and their frequency of occurrence. In our study, we aim to shed light on this issue based on theoretical modelling combined with machine learning (ML) techniques.

The paper is organised as follows. Section~\ref{sect:properties of TEOs} provides a concise characterisation of TEOs and specifies the purpose of our paper. In Sect.~\ref{sect:methods} we present a detailed description of the adopted methodology, including the assumptions made and the software used to generate the theoretical models. We then analyse the obtained models and present our findings in Sect.~\ref{sect:results}. Finally, we summarise the entire work and draw several conclusions in Sect.~\ref{sect:summary-and-conclusions}.

\section{Properties of TEOs and the purpose of the paper}\label{sect:properties of TEOs}
TEOs are tidally forced eigenmodes of a star with frequencies, $\sigma_{nlm}$ (in the co-rotating frame of the star), coinciding with integer multiples, $N$, of the orbital frequency, $f_{\rm orb}$\footnote{We denote the corresponding orbital period as $P_{\rm orb}=1/f_{\rm orb}$.}. The resonance condition can be written as follows:
\begin{equation}\label{equ:resonance-condition}
    f_{Nm}\equiv Nf_{\rm orb} - mf_{\rm s}\approx \sigma_{nlm},
\end{equation}
where $f_{Nm}$ corresponds to the frequency of the tidal forcing in the rotating frame, $f_{\rm s}$ stands for the rotational frequency of the star, while the subscripts $n,l$ and $m$ denote the radial order, degree, and azimuthal order of the specific eigenmode, respectively. This property of TEOs makes them relatively easy to distinguish from other types of pulsations (e.g.,~self-excited oscillations) in frequency spectra, provided the orbital period is known. There are numerous examples of photometrically or spectroscopically detected TEOs \citep[e.g.,][]{2002MNRAS.333..262H,2011ApJS..197....4W,2013MNRAS.434..925H,2017MNRAS.472L..25F,2019ApJ...885...46G,2020ApJ...888...95G,2022ApJ...928..135W}, also in massive binary systems \citep[e.g.,][]{2002A&A...384..441W,2017MNRAS.467.2494P,2021A&A...647A..12K,2022A&A...659A..47K}. Most TEOs are damped normal modes, meaning that without constant tidal forcing they would not be observed in the star. More importantly, because of their damped nature, TEOs dissipate the total orbital energy making the system tighter, more circular, and synchronised with time. On the other hand, if the TEO is naturally an overstable mode it can transfer thermal energy from the star to the orbit via so-called `inverse tides' \citep{2021MNRAS.501..483F}. Regardless of the type of TEOs, they undeniably contribute to the evolution of the (massive) binary system, and can therefore influence the characteristics of the phenomena and objects mentioned above. The efficiency of energy transfer between the stellar interior and the orbit due to TEOs strongly depends on the temporal behaviour of the resonance condition given by Eq.~(\ref{equ:resonance-condition}). It is to be expected that most TEOs are `chance resonances', i.e. resonances in which the aforementioned condition is satisfied for a relatively short time. Under such circumstances, TEOs do not have enough time to reach high amplitudes, hence their ability to dissipate orbital energy is somewhat limited. However, if, after reaching resonance, both frequencies on the left and right sides of Eq.~(\ref{equ:resonance-condition}) evolve at the same rate and direction, TEOs can `tidally lock' for a longer time compared to the chance resonance scenario \citep{2017MNRAS.472.1538F}. This unique variety of TEOs is suspected to be responsible for occasional periods of rapid evolution of the orbital parameters in binary systems \citep{2017MNRAS.472L..25F}.

TEOs are are not only restricted to MS stars, they can also occur in binaries with pre-MS stars \citep{2021AJ....161..263Z}, some compact objects \citep[white dwarfs,][]{2021MNRAS.501.1836Y}, planetary systems \citep{2021ApJ...918...16M} and even planet-moon systems \citep[e.g.,~in the Saturn-Titan system,][]{2020NatAs...4.1053L}.

Although the literature on theoretical studies of TEOs is indeed extensive \citep[see e.g.,][for recent reviews]{2017MNRAS.472.1538F,2021FrASS...8...67G}, the question of their rate of occurrence and the role they play in the evolution of massive stars is still a matter of debate. Unfortunately, only a small number of papers refer exclusively to massive EEVs. \cite{1999A&A...341..842W,1999A&A...350..129W} studied gravity- ($g$) and Rossby-mode TEOs in an uniformly rotating 10$\,M_\odot$ MS star using their own two-dimensional (2D) code for different stellar rotation rates and several orbital configurations. They found that dynamical tides can effectively circularise and tighten the orbits of EEVs in just a few Myrs if resonance locking occurs. However, these and many other previous works on TEOs were done under the assumption of a compact (point-like) secondary companion that is not subject to tidal perturbations during each periastron passage. This is obviously not the case in real binary systems, where both components are responsible for the tidal evolution of the orbit. As theoretically shown by \cite{2001A&A...366..840W}, for an eccentric binary system consisting of two 10\,$M_\odot$ stars, tidal dissipation can be further enhanced due to the simultaneous excitation of tidally-locked TEOs in both components. In spite of the advanced mathematical formalism, the aforementioned papers only dealt with a few assumed component masses and sets of orbital parameters. Only \cite{2003MNRAS.346..968W} attempted a qualitative analysis of the hyperspace of the orbital parameters favouring excitation of TEOs in massive EEVs on MS. He found that for a mass range of 2\,--\,20\,$M_\odot$, the favourable orbital period interval lies between $\sim$5 and $\sim$12\,d when both components are zero-age MS (ZAMS) stars. This interval shifts towards longer orbital periods (up to $\sim$70\,d) for components approaching terminal-age MS (TAMS).

Although, as argued above, the role of TEOs in the life of massive binary systems is still not well understood, we are not aware of any published work that develops the qualitative analysis carried out by \cite{2003MNRAS.346..968W} based on state-of-the-art stellar evolution and oscillations codes. We would like to fill this gap by combining the Modules for Experiments in Stellar Astrophysics\footnote{\url{https://docs.mesastar.org/en/latest/}} \citep[\texttt{MESA},][]{Paxton2011, Paxton2013, Paxton2015, Paxton2018, Paxton2019} stellar structure and evolution code with the \texttt{GYRE}\footnote{\url{https://gyre.readthedocs.io/en/stable/}} \citep{Townsend2013, Townsend2018} non-adiabatic stellar oscillation code. Our study aims to answer the following three questions:
\begin{enumerate}
    \item How many resonances (given by Eq.~(\ref{equ:resonance-condition})) can EEVs experience during their lifetime between ZAMS and TAMS? How does this picture change with different initial parameters of the binary system?
    \item Can we distinguish several distinct types of EEV resonance histories that are statistically related to the initial physical and orbital parameters of binary systems?
    \item Does the resonance history correlate in any way with the properties of EEV near TAMS? For instance, are systems that undergo mass transfer before reaching TAMS also systems with a higher total number of resonances encountered?
\end{enumerate}
In order to answer the last two questions, we use of ML techniques by performing a Uniform Manifold Approximation and Projection\footnote{\url{https://umap-learn.readthedocs.io/en/latest/}} \citep[UMAP,][]{2018arXiv180203426M} dimension reduction analysis of the resonance histories obtained for simulated binary systems.

%-------------------------------------------------------------------
\section{Methods}\label{sect:methods}
Assuming that the dynamical tide excited in the component is dominated by a single TEO close to resonance with the orbital harmonic $N$, one can express its amplitude of luminosity change as proportional to \citep[after][his eq.~2]{2017MNRAS.472.1538F}
\begin{equation}\label{eqn:amplitude}
    A_N\propto q\left(\frac{R}{a}\right)^{l+1}|Q_{nlm}|F_{Nm}\mathcal{L}_N,
\end{equation}
where $q$ is the ratio of the masses of the two components, $R$ stands for the radius of the component in which the TEO is excited, and $a$ is the semi-major axis of the relative orbit. The quantity denoted $Q_{nlm}$ is known as the so-called overlap integral, which describes the spatial coupling between a given oscillation mode and the actual geometry of the tidal potential \citep[][his eq.~4]{2017MNRAS.472.1538F}. In general, the larger the value of $|n|$\footnote{We use $|n|$ instead of $n$ because \texttt{GYRE} assigns negative values of $n$ to $g$ modes and positive ones to $p$ modes.}, the smaller the $Q_{nlm}$, hence eigenmodes with a large number of nodes in the radial direction have a much lower probability of tidal excitation\footnote{More precisely, $Q_{nlm}$ does not vary strictly monotonic with $n$ and can change significantly between consecutive modes for given $l$ and $m$. However, the overall trend of $Q_{nlm}$ peaks for low values of $|n|$ and falls sharply for $|n|\gg0$. For a more detailed discussion on the behaviour of $Q_{nlm}$ see, e.g., \cite{2012MNRAS.421..983B}.}. In addition to $Q_{nlm}$, the Hansen coefficient $F_{Nm}$ \citep[][his eq.~5]{2017MNRAS.472.1538F} is responsible for the temporal coupling of the forced normal mode and the $N^{\rm th}$ component in the Fourier expansion of the orbital motion. Quantitatively, it expresses the intuitive principle that for more eccentric orbits, TEOs with larger orbital harmonic numbers will be excited. This is because, as the eccentricity increases, the periastron passage takes less time for a given orbital period, so eigenmodes with higher frequencies better `match' rapidly changing gravitational field, in terms of time scale. Nevertheless, for very high $N$, the $F_{Nm}$ drops rapidly (almost exponentially). This particular property of $F_{Nm}$ is responsible for the lack of excitation of the TEOs with extremely high $N$. It is clear here that the frequency range of TEOs in massive and intermediate-mass MS stars is limited on two sides independently by $Q_{nlm}$ and $F_{Nm}$. On the low-frequency side, $Q_{nlm}$ prevents the excitation of $g$ modes with very high $|n|$, while on the high-frequency side $F_{Nm}$ decreases sharply, strongly limiting the possible excitation of pressure ($p$) modes characterised by high radial orders. The last term in Eq.~(\ref{eqn:amplitude}), i.e. $\mathcal{L}_N$, denotes the detuning factor given by the following formula,
\begin{equation}\label{eq:L_N}
    \mathcal{L}_N= \frac{f_{Nm}}{\sqrt{(\sigma_{nlm}-f_{Nm})^2+\gamma_{nlm}^2}},
\end{equation}
where $\gamma_{nlm}$ stands for damping/growth rate of the normal mode. This Lorentzian-like factor reflects the mismatch between $f_{Nm}$ and $\sigma_{nlm}$. Given the typical values of $|\gamma_{nlm}|$ for $g$ and $p$ modes in massive and intermediate-mass MS stars (of the order of $\sim10^{-7}$\,--\,$10^{-3}$\,\cd), $\mathcal{L}_N$ is extremely sensitive to the difference $(\sigma_{nlm}-f_{Nm})$. Hence, among many other factors, the $\mathcal{L}_N$ undoubtedly plays a key role in the excitation of TEOs.

While the precise prediction of TEO amplitude is a difficult task\footnote{In order to reliably predict the photometric amplitude of a TEO, one needs to: (1) determine the exact value of $\mathcal{L}_N$, which is almost impossible given the uncertainties in both observations and stellar models, (2) calculate the corresponding $Q_{nlm}$ and (3) know the eigenfunction of luminosity fluctuations at the photospheric level, which is a challenge for radiation pressure-dominated atmospheres of early-type stars (with intense stellar winds). In addition, the equilibrium amplitude of a linearly driven TEO is determined by various non-linear effects, for instance by multi-mode coupling \citep{2020ApJ...896..161G,2022MNRAS.517..437G}.}, we are interested in analysing the changes in resonance conditions dictated by the sum of all the contributing detuning factors with passing time, $t$. Let us define the following dimensionless quantity,
\begin{equation}\label{eqn:L(t)}
    \mathcal{L}(t)\equiv \sum\limits_{nlm} \sum\limits_{N=1}^{N_{\rm max}} \mathcal{L}_N(t).
\end{equation}
In contrast to $\mathcal{L}_N$, associated with a single orbital harmonic, $\mathcal{L}$ reflects the overall chance of TEOs being excited in the EEV component. However, we must stress at this point that it does not carry direct information on the amplitude of potential TEOs. The first summation in Eq.~(\ref{eqn:L(t)}) applies to all the normal modes we consider in the modelling (Sect.~\ref{sect:asteroseismic-calculations}). Obviously, the second summation in Eq.~(\ref{eqn:L(t)}) should run from $N=1$ to $+\infty$, but due to time and physical constraints one has to truncate the series at some reasonably chosen $N_{\rm max}$. For a detailed description of the selection of $N_{\rm max}$ values see Sect.~\ref{sect:asteroseismic-calculations}.

In order to try to answer the questions raised in Sect.~\ref{sect:properties of TEOs}, we have synthesised 20,000 binary evolution models and calculated $\mathcal{L}(t)$ for both components in each of them. The whole procedure is described extensively in the next four subsections.

\subsection{General assumptions}\label{sect:methods:general assumptions}
From a practical point of view, a fully consistent calculation of the evolution of binary systems taking TEOs into account is very time-consuming, as it requires time steps shorter than the times at which the resonances occur (several orders of magnitude shorter than the nuclear time scale, cf.~Fig.~\ref{fig:resonance-curves-sample}). It would take an enormous amount of time to perform such consistent calculations for 20,000 binaries with hundreds of resonances occurring in each of them. Therefore, to make our project both feasible and still scientifically useful, the models were synthesised under the general assumption that each resonance encountered by the EEV components is a chance resonance. By sacrificing the ability to track resonantly-locked TEOs, we are able to decouple evolutionary and seismic calculations and run them independently, greatly simplifying the whole problem. We believe that we can to do this for three reasons: (1) we are only interested in obtaining some general statistical information about the resonance conditions in a large number of simulated binary systems, (2) the phenomenon of resonance locking is rare compared to the rate of chance-resonance events, and (3) the impact of chance-resonance TEOs on the orbit is limited due to their relatively short time of existence \citep[e.g.,][]{1999A&A...350..129W}.
In conclusion, we focus on finding candidate binaries that may or may not experience numerous TEOs, rather than precisely predicting their actual evolutionary histories, which is beyond the scope of this paper. We believe that our results will serve as a starting point for more detailed calculations performed for the most interesting cases of massive EEVs.

\subsection{Synthesis of binary evolution models}\label{sect:methods-Synthesis of binary evolution models}
Since we assumed that we could separate stellar and orbital evolution from seismic calculations, we first generated a set of binary evolutionary tracks and only then performed seismic analysis on them to find $\mathcal{L}(t)$.

\subsubsection{Initialisation of models}\label{sect:methods-initialization of models}
We used the latest open-source 1D stellar evolution code \texttt{MESA} (release~15140) compiled with the \texttt{MESA} Software Development Kit \citep[version~21.4.1,][]{2021zndo...2603136T} to compute a set of 20,000 binary evolution models. The \texttt{MESAbinary} module \citep{Paxton2015} allows the simultaneous evolution of binary system components and their orbits. Throughout this paper, we use the subscripts `1' and `2' to denote the primary (initially more massive) and secondary components, respectively. 

\begin{figure}
\centering
\includegraphics[width=\hsize]{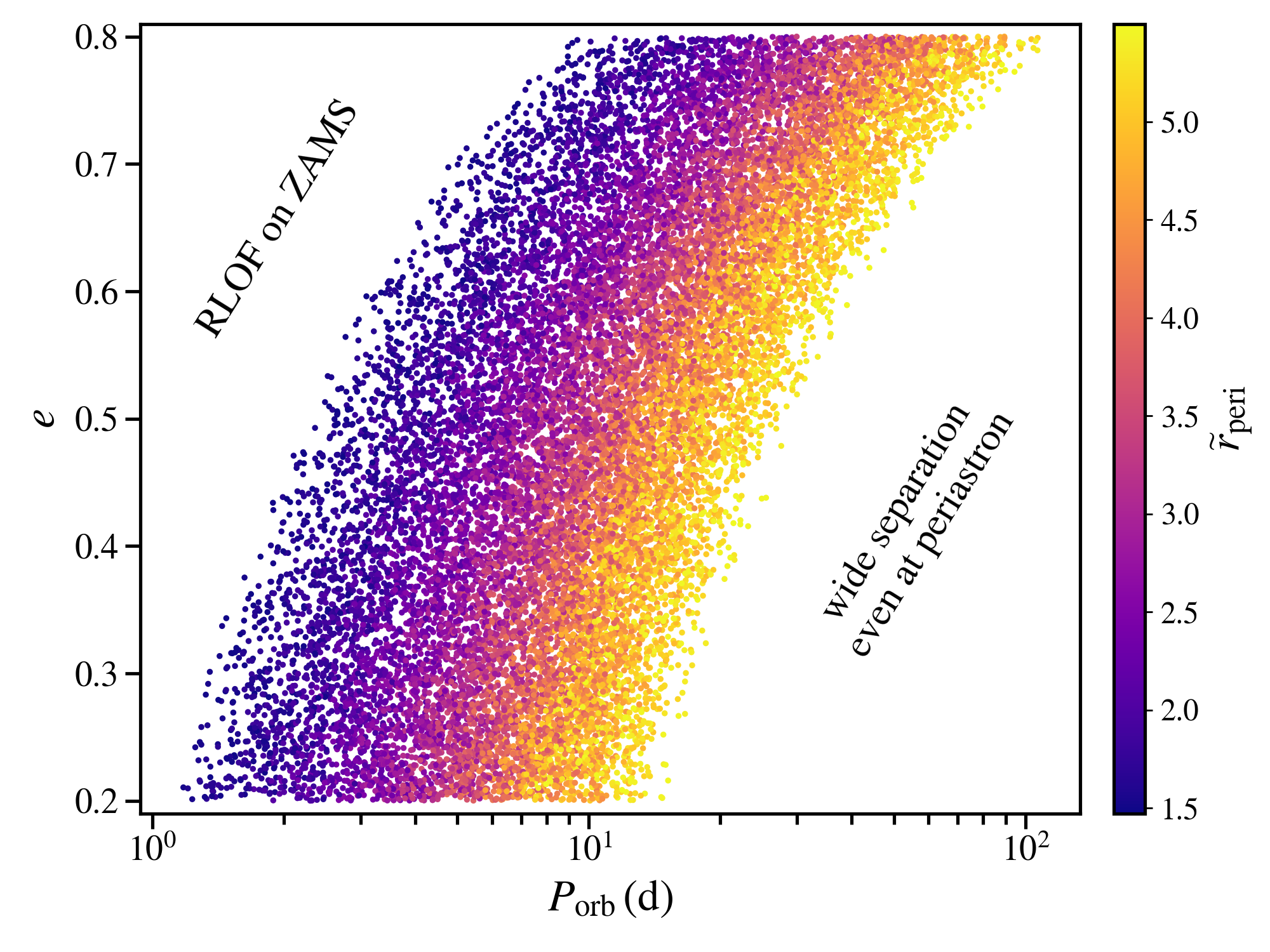}
\caption{Distribution of initial orbital period and eccentricity for a sample of 20,000 binaries evolved in our project. The initial normalised separation at the periastron is colour-coded. The upper left-hand corner corresponds to the ZAMS EEVs, which experience RLOF at the periastron and should therefore rapidly circularise their orbits. The lower right-hand corner, on the other hand, is where relatively widely-separated binaries can be found.}
\label{fig:P-e-ini-plane}
\end{figure}

We assumed that both components have the same chemical composition with metallicity $Z=0.02$ and a solar-scaled mixture of elements taken from \cite{1998SSRv...85..161G}. Since we were only interested in massive and intermediate-mass MS EEVs that can exhibit TEOs during their lifetime, the initial systems consisted of two stars lying on the ZAMS and were characterised by parameters randomly drawn from the following uniform distributions, $\mathcal{U}_{[\alpha,\beta]}$, on specific intervals $[\alpha,\beta]$.
\begin{itemize}
    \item Mass of primary component, $\log(M_1/M_\odot)\sim  \mathcal{U}_{[\log 5,\log 30]}$. A uniform distribution on a logarithmic scale was used instead of a linear scale to cover the Hertzsprung-Russell diagram (HRD) with more evenly distributed evolutionary tracks.
    \item The mass ratio, $q\equiv M_2/M_1\sim \mathcal{U}_{[0.2,0.95]}$, where $M_2$ corresponds to the mass of the secondary component. The lower limit for $q$ was introduced due to the fact that the efficiency in driving TEOs scales with $q$ (cf.~Eq.~(\ref{eqn:amplitude})), so it is less likely to observe TEOs in a binary system at a small value of the mass ratio. Moreover, if the generated $q$ corresponded to $M_2<2M_\odot$, a redraw was performed.
    \item Eccentricity, $e\sim \mathcal{U}_{[0.2,0.8]}$. Range typical of the observed EEVs.
    \item Periastron distance, $r_{\rm peri}$, normalised to the sum of components' radii, $\widetilde{r}_{\rm peri}\equiv r_{\rm peri}/(R_1+R_2)\sim \mathcal{U}_{[1,5.5]}$. However, if the generated system was initially Roche-lobe overflowing (RLOF) at the periastron, a redraw was performed. We also assumed an upper value of $\widetilde{r}_{\rm peri}$ because the overall strength of tidal forces decays as $r_{\rm peri}^{-3}$ and simulating widely-separated systems would contradict the aim of this paper.
    \item Tidally-enhanced wind factor, $B_{\rm wind}\sim \mathcal{U}_{[32,896]}$. Introduced by \cite{1988MNRAS.231..823T} for red giants residing in binary systems, it accounts for the tidal enhancement of the stellar wind mass-loss rate due to the presence of a nearby companion. The \emph{ad hoc} chosen range of $B_{\rm wind}$ corresponds to a maximum amplification of the `nominal' wind mass-loss rate by a factor of 1.5\,--\,10 \citep[cf.][their eq.~2]{1988MNRAS.231..823T}.
    \item The angular rotational velocity divided by its critical value\footnote{By critical rotational velocity we mean the situation when the effective gravity at the stellar equator is zero, i.e.~the centrifugal force and the Eddington factor, $\Gamma$, balance the true gravity. \texttt{MESA} estimates this quantity as $\Omega_{\rm crit}=\sqrt{(1-\Gamma)GM/R^3}$, where $G$ is the gravitational constant and $\Gamma\equiv L_{\rm rad}/L_{\rm Edd}$. The $L_{\rm rad}$ and $L_{\rm Edd}$ denote the radiative luminosity and Eddington luminosity of the star, respectively.}, $\Omega/\Omega_{\rm crit}\sim \mathcal{U}_{[0.1,0.5]}$. The assumed range of initial $\Omega/\Omega_{\rm crit}$ translates into the linear equatorial velocities between $\sim$50\,km/s and $\sim$320\,km/s in our simulations and reflects the significant non-zero rotation velocities observed in massive young MS stars \citep[e.g.,][]{2006A&A...457..265D,2008A&A...479..541H}. 
    \item The overshoot mixing parameter, $f_{\rm ov}\sim \mathcal{U}_{[0.015,0.025]}$. In our calculations, the overshooting of the material above the convective, hydrogen-burning core was treated in the exponential diffusion formalism developed by \cite{2000A&A...360..952H}. An adjustable parameter, $f_{\rm ov}$, describes the spatial extent of the overshoot layer in terms of the local pressure-scale height, but its value for massive stars is still under debate. We adopted the range of $f_{\rm ov}$ after \cite{2017ApJ...835..290O}.
\end{itemize}
The parameters presented above were generated independently for each EEV system. Moreover, the last two parameters, $\Omega/\Omega_{\rm crit}$ and $f_{\rm ov}$, were drawn independently for each component, so the final hyperspace of parameters explored in our simulations included $\{M_1,q,e,\widetilde{r}_{\rm peri},\Omega/\Omega_{\rm crit,1},\Omega/\Omega_{\rm crit,2},f_{\rm ov,1},f_{\rm ov,2},B_{\rm wind}\}$. Figure~\ref{fig:P-e-ini-plane} shows our initial sample of generated EEVs in the orbital period versus eccentricity diagram. As expected, they occupy the upper envelope of the aforementioned plane with the upper boundary dictated by the onset of periastron RLOF on ZAMS. The rest of the necessary parameters and `physics switches' were identical for each simulated binary. We will now briefly describe them below.

\subsubsection{Integration of the evolution}
Nuclear reaction rates were calculated using `\texttt{basic.net}' option in \texttt{MESA}. We used a convective premixing scheme \citep[][their Sect.~5]{Paxton2019} in combination with the Ledoux criterion to define the boundaries of convective instability. This specific approach of treating convection agrees with the results of modern 3D hydrodynamic simulations \citep{2022arXiv220306186A}. Convective mixing was incorporated into the models via mixing length theory (MLT) in the `Cox' formalism \citep[][their chap.~14]{1968pss..book.....C} with the value of the solar-calibrated mixing length parameter $\alpha_{\rm MLT}=1.82$\footnote{There is some evidence that $\alpha_{\rm MLT}$ may depend on global stellar parameters such as mass \citep{2006MNRAS.368.1941Y} or metallicity \citep{2018ApJ...858...28V}. It is also very likely that $\alpha_{\rm MLT}$ is sensitive to the evolutionary stage of the star and the type of convection zone \citep[e.g.,][]{2015A&A...577A.134W}. Here, we have assumed a constant value of $\alpha_{\rm MLT}$ for simplicity.} \citep{2016ApJ...823..102C}. As mentioned earlier, exponential overshoot mixing above the convective core was also included\footnote{Similarly to the $\alpha_{\rm MLT}$, $f_{\rm ov}$ also can depend on different stellar parameters \citep[e.g.,][]{2014A&A...570L..13C}.}, but we neglected overshooting in the non-burning convection zones. For stars with masses $\geqslant 15\,M_\odot$, we activated the treatment of convection as `MLT++' \citep[][their Sect.~7.2]{Paxton2013} to reduce superadiabaticity in convective zones dominated by radiation pressure. Since we used Ledoux criterion, semiconvection could appear in our stars with its efficiency parameter, $\alpha_{\rm sc}=0.01$ \citep{1985A&A...145..179L}. In our case, semiconvection sometimes occurred in chemically-modified layers left by the shrinking core.

Upon initialisation at the ZAMS, we relaxed both components in $\sim$100 steps so that they rotated rigidly. Later, we allowed our stars to rotate differentially during their evolution, according to the so-called shellular approximation of rotation \citep{1997A&A...321..465M}. Throughout the entire evolution, we assumed that the rotation axes of the stars are perpendicular to the orbital plane. \texttt{MESAstar} uses the mathematical formalism of \cite{2000ApJ...528..368H} and \cite{2005ApJ...626..350H} to apply structural corrections, perform different types of rotationally induced mixing and ,,diffusion'' of angular momentum between adjacent shells. The following rotational mixing mechanisms were taken into account in \texttt{MESA}: dynamic shear instability, secular shear instability, Eddington-Sweet circulation, Solberg-H\o iland instability, and Goldreich-Schubert-Fricke instability \citep[all described in detail by][]{2000ApJ...528..368H}. Even the combined mixing coefficients of the aforementioned rotational instabilities can be zero in some parts of the star. However, this is clearly unrealistic due to the presence of a nearby companion that induces additional mixing throughout the star. To at least approximately account for this process, we did not allow the total mixing coefficient, $D_{\rm mix}$, to fall below $10^5$\,cm$^2$/s. This particular arbitrarily-selected value is related to the mixing time scale, $\tau_{\rm mix}\sim (\Delta r)^2/D_{\rm mix}\approx 15$\,Myr at radial distance, $\Delta r= 0.1\,R_\odot$. We cannot conceal here that rotation and mixing profiles in MS stars are still poorly understood (except in the solar case). There are no definitive conclusions as to what mixing mechanisms and whether they actually occur in massive and intermediate-mass MS stars \citep[see e.g.,][for a discussion of this problem and its asteroseismic inference from B-type dwarfs]{2021NatAs...5..715P,2022ApJ...930...94P}.

Mass losses due to the radiation-driven stellar wind were calculated according to the prescription given by \cite{2001A&A...369..574V}. Their formulae take into account the presence of a so-called bi-stability jump around the effective temperature of $T_{\rm eff}\approx$\,26,000\,K, caused by ionization and recombination of some Fe ions. Nevertheless, the presence of a bi-stability jump is still questionable and there is some evidence that the associated almost instantaneous change in the mass-loss rate may not be real \citep[cf.][]{2021A&A...647A..28K,2022arXiv220308218B}. `Nominal' wind mass-loss rates in our simulations were modified in two ways: (1) the rate was amplified by the aforementioned tidal mechanism, parametrized by $B_{\rm wind}$ \citep{1988MNRAS.231..823T} and (2) the effect of fast rotation at the surface, which can amplify the mass-loss rate, was accounted for by the simplified power-law description given by \cite{2000ApJ...528..368H} (their Sect.~2.6). We assumed that the mass loss through the wind is completely non-conservative, i.e.~there is no mechanism that could transfer some material back to the star or to a companion.

As we already mentioned above, \texttt{MESAbinary} allows the simultaneous integration of some stellar and orbital parameters that are coupled to each other in a binary system. We have switched on the \texttt{MESA} controls responsible for changes in the total orbital angular momentum caused by: (1) gravitational wave radiation, (2) wind mass loss and (3) tidal spin-orbit coupling. For the first process, the rate of orbital momentum loss was calculated assuming point masses. The mass loss through the stellar wind was completely non-conservative, so the angular momentum lost via this channel was equal to the angular momentum carried by the wind. The phenomenon (3), contributing to the evolution of eccentricity, orbital and spin angular momenta, was modelled using the theory of tidal interactions for radiative envelopes developed by \cite{1977A&A....57..383Z}, \cite{1981A&A....99..126H} and \cite{2002MNRAS.329..897H}, after being adapted to the shellular approximation of rotation. For stars with radiative envelopes, tidal dissipation processes are dominated by tidally excited gravity modes that propagate to the stellar surface, where they gain relatively large amplitudes and experience effective radiative damping (due to the short local thermal time scale) and nonlinear damping. Consequently, they deposit their energy and angular momentum in the outer layers of the envelope. Following earlier calculations of \cite{1977A&A....57..383Z}, \cite{2002MNRAS.329..897H} delivered convenient formulae to describe the tidal synchronisation and circularisation time scales associated with the aforementioned phenomenon. Using these time scales combined with the formalism presented by \cite{1981A&A....99..126H}, \texttt{MESAbinary} integrates the evolution of the eccentricity and updates spin angular frequency of each shell in the stellar model. Therefore, our calculations in \texttt{MESAbinary} took into account the approximate influence of the dynamical tide on the orbit, at least up to the lowest possible order. Of course, the tidal evolution formalism implemented in \texttt{MESAbinary} does not include the effects of resonance locking. For explicit formulae describing tidal processes in \texttt{MESAbinary}, we refer to \cite{Paxton2015} (their Sect.~2).

We have completely ignored the effects of magnetic fields, while bearing in mind that they may mainly affect the actual stellar wind mass-loss rates, the efficiency of internal mixing processes and synchronisation/circularisation time scales (e.g. via the magnetic braking mechanism). The impact of fossil magnetic fields on the evolution of massive and intermediate-mass stars was recently described by \cite{2022arXiv220906350K}.

All details on the parameters of our models in \texttt{MESA} can be found in Appendix~\ref{appendix:mesa}, where we present the contents of our \texttt{MESA} inlists. A concise description of the micro- and macrophysics data sources used by \texttt{MESA} is provided in Appendix~\ref{appendix:mesa-data}.

\subsubsection{Termination conditions}\label{sect:methods-termination}
The evolution of the binary system was carried out until at least one of the following termination conditions was met for any of the components:
\begin{enumerate}
    \item The component reached TAMS, i.e.~the central mass abundance of hydrogen fell below $X_{\rm c}\leqslant10^{-4}$.
    \item The eccentricity was reduced to $e\leqslant 0.01$.
    \item The rotation velocity reached $\Omega/\Omega_{\rm crit}=0.75$ at the stellar surface.
    \item Episodic mass transfer between components due to the RLOF in the periastron began.
\end{enumerate}

The reasons behind providing the termination conditions outlined above are as follows. Our study is exclusively dedicated to the MS phase of the evolution of EEVs, hence the first condition has to be enforced. The second condition is self-explanatory, since we are interested in non-zero eccentricities that allow for TEO excitation\footnote{In theory, components of circular systems ($e=0$) can also exhibit TEOs, provided they do not rotate synchronously. However, the number of modes observable as TEOs in these systems is much smaller than the number of potential TEOs in EEVs.}. The third condition is related to the convergence problems that can occur in \texttt{MESAbinary} when one of the components nearly approaches the break-up velocity of rotation. Numerous assumptions and descriptions of rotation-related phenomena reach the limits of their applicability in \texttt{MESA} for $\Omega/\Omega_{\rm crit}\approx 1$. Since for $\Omega/\Omega_{\rm crit}\gtrsim 0.75$ the deviation from spherical symmetry becomes significant, a 1D treatment of the problem is no longer adequate. For instance, the way in which such a star loses mass becomes fundamentally different from the isotropic case. We have therefore decided to stop integrations under such circumstances. The last condition is related to the difficulty in correctly describing an episodic (near-periastron) RLOF, when a `blob' of material could be ejected from the RL-filling component during each periastron passage. However, this kind of orbital phase-dependent RLOF is not expected to be observed in a binary for a long time due to strong tidal forces. They should effectively suppress the eccentricity, making the system circular (and so the second condition can be quickly met).

\subsection{Asteroseismic calculations}\label{sect:asteroseismic-calculations}
\begin{figure}
\centering
\includegraphics[width=\hsize]{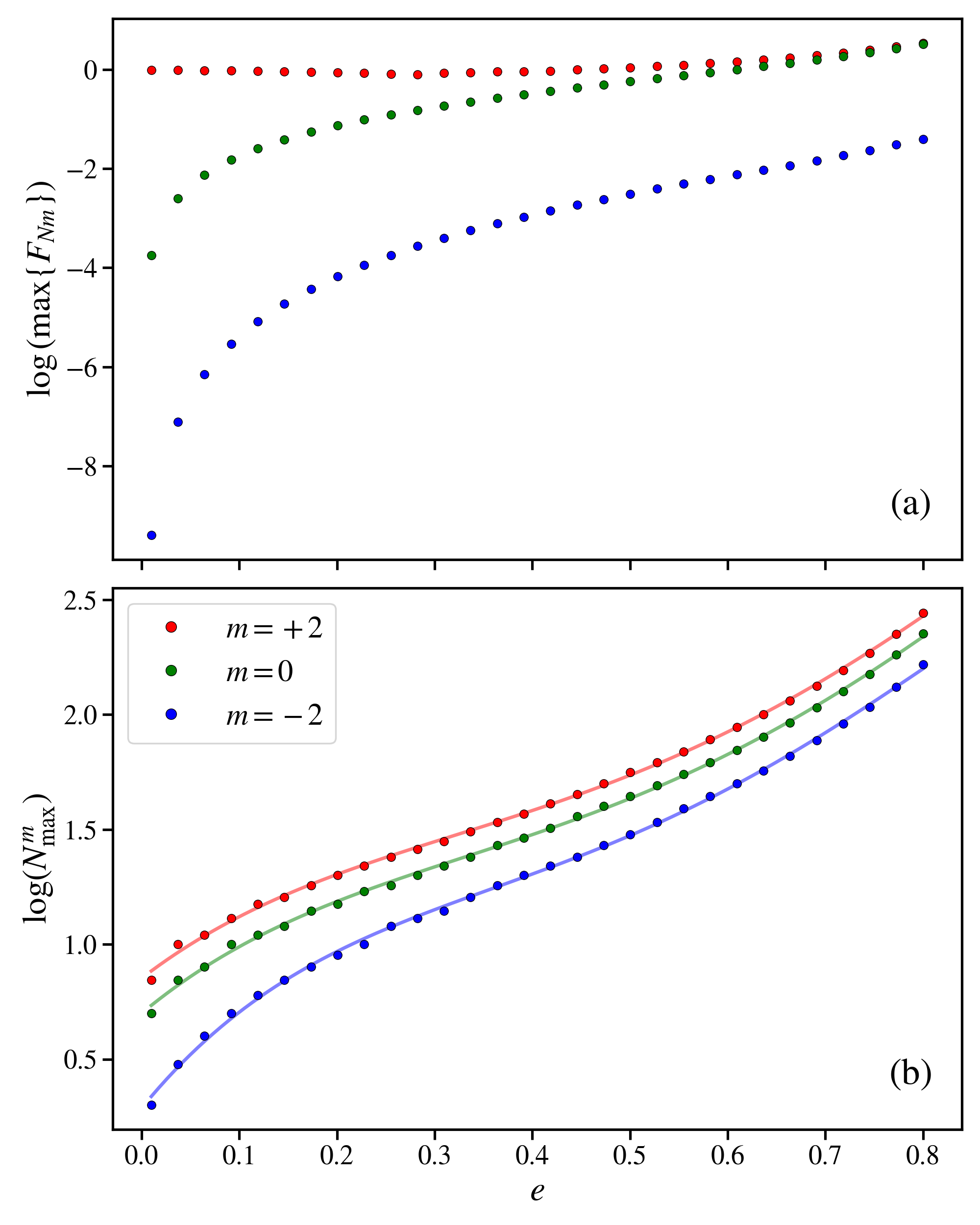}
\caption{(a) Maximum values of the Hansen coefficients $F_{Nm}$ versus eccentricity for $l=2$ modes and three different azimuthal orders, $m=-2,0,+2$ denoted by blue, green, and red points, respectively. We note the marginal contribution of the $m=-2$ modes; (b) Dependence of $\log N_{\rm max}^m$ on eccentricity with the same colour-coding as in panel a. The colour solid lines represent the best fits of the fourth-degree polynomials, which we used to determine frequency ranges in the asteroseismic calculations.}
\label{fig:hansen-Nmax-fit}
\end{figure}

A consequence of the assumption of the aligned vectors of the orbital and spin angular momenta is a rule for selecting the geometry of modes that can be tidally excited. Under such conditions, a normal mode can be tidally excited only if $\mod(|l+m|,2)=0$, e.g. the $l=2$ TEOs will be characterised only by $m=-2,0,+2$. Here we restrict our study to only $l=2$, $m=0,+2$ modes because of two reasons. First, $l=2$ modes correspond to the dominant component in the series expansion of the variable tidal potential. Modes with $l>2$ undergo much weaker excitation due to the dependence on $(R/a)^{l+1}$, which enters Eq.~(\ref{eqn:amplitude}). Second, the values of $F_{N,-2}$ are very small compared to their $m=0,+2$ counterparts. This can be easily seen in Fig.~\ref{fig:hansen-Nmax-fit}a, where we have plotted the maximum values of $F_{Nm}$ for $m=-2,0,+2$ and different eccentricities. $F_{N,-2}$ is approximately at least 2\,--\,3 orders of magnitude smaller than $F_{N,0}$ or $F_{N,2}$.

For each model of the stellar internal structure that was saved during the synthesis of binaries in \texttt{MESA}, we calculated the oscillation spectrum using the \texttt{GYRE} code in the non-adiabatic regime and the second-order Gauss-Legendre Magnus integrator. The frequencies $\sigma_{n,2,0}$ and $\sigma_{n,2,+2}$ corresponding to the non-adiabatic calculations were searched by \texttt{GYRE} based on the preliminary adiabatic calculations. Rotational effects (Coriolis force) were taken into account using the so-called traditional approximation of rotation \citep[e.g.,][their Sect.~3.8]{2010aste.book.....A}. We searched for eigenvalues in the family of gravito-acoustic solutions. We assumed the necessary (differential) rotation profile inside the star from the \texttt{MESA} model.

As we argued in Sect.~\ref{sect:methods}, it is necessary to choose a reasonable range of frequencies to scan for eigenvalues based on $Q_{nlm}$ and $F_{Nm}$. Therefore, we only searched for modes with $|n|\leqslant 30$ and frequencies, $\sigma_{n,2,0}\in (f_{\rm orb},N_{\rm max}^{m=0}f_{\rm orb})$ or $\sigma_{n,2,+2}\in (\max \{0,f_{\rm orb}-2f_{\rm s, core}\},N_{\rm max}^{m=+2}f_{\rm orb}-2f_{\rm s, core})$. In the ranges shown, $N_{\rm max}^{m=0}$ and $N_{\rm max}^{m=+2}$ refer to the limits of $N$ due to the decrease in $F_{N,0}$ and $F_{N,2}$, respectively. The $f_{\rm s, core}$ is the core rotation frequency. We defined $N_{\rm max}^{m=0}$ and $N_{\rm max}^{m=+2}$ as $N$ for which $F_{N,0}$ or $F_{N,2}$ is equal to $10^{-8}$, i.e.~$F_{Nm}$ starts to effectively prevent excitation of TEOs. In practice, we numerically calculated the $F_{Nm}$ functions\footnote{Using eq.~5 presented by \cite{2017MNRAS.472.1538F}.} for different eccentricities and obtained the $\log N_{\rm max}^m(e)$ relations as a fit of a fourth-degree polynomial to a set of its discrete points. A summary of this process is shown in Fig.~\ref{fig:hansen-Nmax-fit}b. For low-$e$ orbits, the typical range of $N$ favourable for the excitation of TEOs reaches $N\sim10^1$, in contrast to highly eccentric orbits, which may exhibit as much as $N\approx100$\,--\,200 TEOs. Figure~\ref{fig:hansen-Nmax-fit}b also shows another feature of $m=-2$ modes that makes them inferior candidates for TEOs compared to axisymmetric and prograde modes -- as potential TEOs they always span a narrower range of orbital harmonic numbers.

Defining the frequency range for $\sigma_{n,2,0}$ is quite straightforward, as these are axisymmetric modes and their frequencies do not change when switching between inertial and co-rotating frames. The situation is quite different when it comes to the $m=+2$ modes. This time, due to the differential rotation inside the star, $\sigma_{nlm}=\sigma_{nlm}(r)=\overline{\sigma}_{nlm}-mf_{\rm s}(r)$, where $r$ is the radial coordinate in the stellar interior and $\overline{\sigma}_{nlm}$ is oscillation frequency in the inertial frame. For some eigenmodes, $\sigma_{nlm}$ may change its sign somewhere in the star, depending on the shape of the rotational profile. This location is known as the critical layer, where $\sigma_{nlm}(r)\rightarrow 0$, and such a mode experiences severe damping due to its very short radial wavelength \citep[e.g.,][]{2013A&A...553A..86A}. To exclude these modes from our experiment, the maximum frequency of $\sigma_{n,2,+2}$ was set to $(N_{\rm max}^{m=+2}f_{\rm orb}-2f_{\rm s, core})$\footnote{We note that this frequency is expressed in a rest frame co-rotating with the stellar core.}. This is because during evolution the core rotates almost rigidly and faster than the envelope, hence the difference $(N_{\rm max}^{m=+2}f_{\rm orb}-2f_{\rm s, core})\leqslant(N_{\rm max}^{m=+2}f_{\rm orb}-2f_{\rm s, env})$, where $f_{\rm s, env}$ stands for the rotation frequency of the outermost part of the envelope.

More details of our calculations performed in \texttt{GYRE} can be found in Appendix~\ref{appendix:gyre}, where we present the explicit contents of our \texttt{GYRE} input file.

\subsection{Derivation of $\mathcal{L}(t)$}\label{sect:methods-derivation-of-L}
The introduction of differential rotation also has consequences when it comes to interpreting the resonance condition from Eq.~(\ref{equ:resonance-condition}). The quantity $f_{\rm s}$ is no longer a constant value, so one has to decide which $f_{\rm s}$ to choose. Theoretical studies imply the induction of $g$-mode TEOs (especially of high radial order) primarily near the convective core boundary \citep[e.g.,][]{1989ApJ...342.1079G} for stars with radiative envelopes. However, the resonances in our simulations are also due to $p$ or $g$ modes with small radial order. Therefore, we decided to apply our resonance condition to the envelope\footnote{In the exact approach, different resonance conditions would have to be used for different modes, depending on the radial coordinate inside the star where a given TEO is dominantly excited. Here we assume a single form of resonance condition for all modes.} (not to the interface region near the core boundary), rewriting Eq.~(\ref{equ:resonance-condition}) more accurately as
\begin{equation}\label{equ:resonance-condition-rotation}
    f_{Nm}\equiv Nf_{\rm orb} - mf_{\rm s, env}\approx \sigma_{nlm},
\end{equation}
and use it in the subsequent modelling of $\mathcal{L}(t)$. It is essential to note at this point that the resonance condition given by Eq.~(\ref{equ:resonance-condition-rotation}) refers to $f_{Nm}$ and $\sigma_{nlm}$ expressed in a frame co-rotating with the outer stellar envelope. Although in principle the morphology of $\mathcal{L}(t)$ depends on the choice of the specific resonance condition, we note that it does not affect at all resonances due to $m=0$ modes and should not significantly affect resonances corresponding to $p$ modes or low-$|n|$, $m=+2$ $g$ modes.  

\begin{figure*}
   \centering
   \includegraphics[width=0.9\hsize]{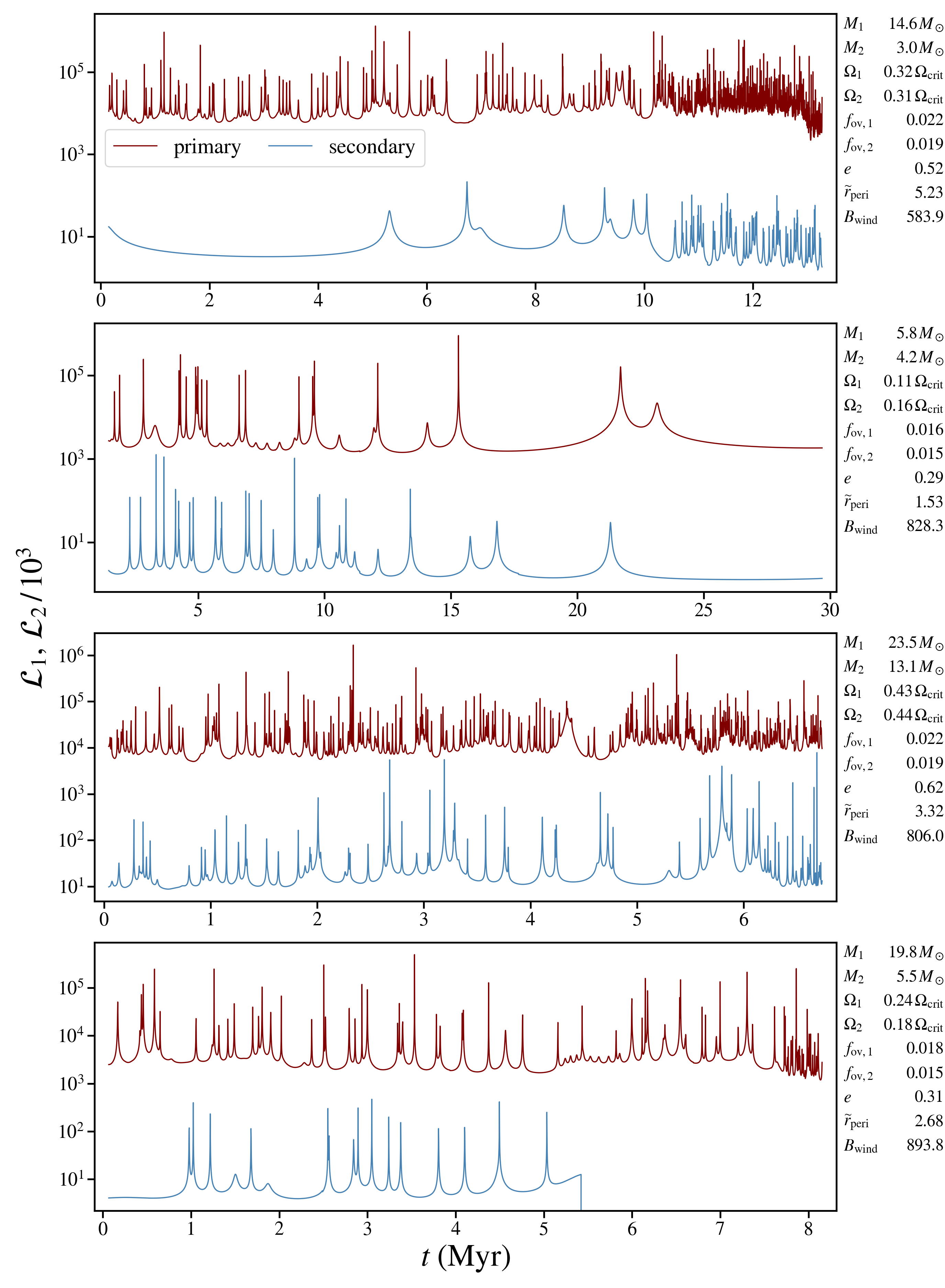}
   \caption{Sample of resonance curves obtained as described in Sect.~\ref{sect:methods}. Each panel corresponds to a different arbitrarily-chosen binary system with the rounded values of their initial parameters given on the right. The dark red and blue curves reflect the behaviour of $\mathcal{L}(t)$ for the primary and secondary components, respectively. For clarity, $\mathcal{L}_2(t)$ has been shifted vertically by three orders of magnitude downwards. Time $t=0$ indicates ZAMS. A sudden break in $\mathcal{L}_2(t)$ on the bottom panel (after about 5.5\,Myr) indicates $\mathcal{L}_2=0$, i.e.~the absence of any resonances. The differences in the height of the peaks are due to different values of $\gamma_{nlm}$ and $\min\{|\sigma_{nlm}-f_{Nm}|\}$ for excited TEOs.}
   \label{fig:resonance-curves-sample}
\end{figure*}

Having a set of eigenfrequencies calculated by \texttt{GYRE} and knowing the history of the binary evolution from \texttt{MESA}, we performed the summation shown in Eq.~(\ref{eqn:L(t)}). However, this was not a direct summation running over the models saved by \texttt{MESA} and \texttt{GYRE}, as their temporal resolution was still too coarse compared to the duration of a typical resonance. To circumvent this problem, we interpolated the temporal variations of each oscillation frequency and all necessary parameters of the binary system using Akima cubic spline functions \citep{10.1145/321607.321609}. Then, we were able to calculate the values of $\mathcal{L}(t)$ on a uniformly-spaced time grid with a constant time step of 2,000 years, which we assumed to be identical for all EEVs in our simulations. From here on, we will use the term `resonance curve' as a proxy for the $\mathcal{L}(t)$ time series. Figure~\ref{fig:resonance-curves-sample} shows a compilation of example resonance curves, although we postpone discussion of these to Sect.~\ref{sect:results}. Together with the initial parameters of binary systems, resonance curves are particularly important to us in this study.

\subsection{ML analysis of the resonance curves}\label{sect:methods-ML analysis of the resonance curves}
Although in Sects.~\ref{sect:results-total-number-of-resonances} and \ref{sect:results-Distribution of resonances over time} we analyse the resonance curves based on various statistics, due to their global nature we do not distinguish many details that are `hidden' in the resonance curves. To characterise the morphology of all resonance curves in more detail (without having to perform a visual classification, which is almost impossible due to the number and complexity of the data set), we applied dimensionality reduction methods. With these, we were able to explore the topology spanned by the morphological features of the resonance curves. We carried out the entire analysis described here separately for the sets of curves $\mathcal{L}_1(t)$ and $\mathcal{L}_2(t)$, corresponding to the primary and secondary components, respectively.

As a first step, we summarised each resonance curve with a vector $\vec{Q}$ that described its morphological features. We focused our attention on two particular features: (1) the distribution of $\log(\mathcal{L})$ values and (2) the distribution of apparent maxima at a normalised time, $t/T_{\rm max}$, where $T_{\rm max}$ stands for the $\max\{t\}$. In practice, we calculated vectors $\vec{Q}_{x}$ and $\vec{Q}_{y}$ which contained sets of 1,000 quantiles of normalised times corresponding to local maxima of $\mathcal{L}(t)$ and 1,000 quantiles of $\log(\mathcal{L})$, respectively. The levels of both calculated quantiles were spanned evenly from 0 to 1. $\vec{Q}_x$ describes the overall distribution of apparent maxima in time, reporting changes in the rate of resonance occurrence. We deliberately used normalisation by $T_{\rm max}$ because we want the results to be sensitive only to the relative distribution of the resonance events over the lifetime of the EEV. Otherwise, its values would be strongly correlated with the length of the resonance curve itself\footnote{Which in turn is an almost a direct approximation for the mass of the primary component.}, rather than with the distribution of resonances over time. On the other hand, $\vec{Q}_y$ encapsulated the combined information about the mean level of $\log(\mathcal{L})$, any long-term trends in the resonance curve and the distribution of the heights of the maxima. In contrast to the $\vec{Q}_x$, we did not apply any normalisation to $\vec{Q}_y$ as its absolute values carry valuable information about the strength of the resonances and the average level of the entire resonance curve. The final vector $\vec{Q}$ was constructed as the concatenation of $\vec{Q}_x$ and $\vec{Q}_y$, which had previously been scaled using the variance in the sets of all $\vec{Q}_x$ and $\vec{Q}_y$. The resulting $\vec{Q}$ has a total of 2,000 dimensions.

In the next step, we performed a preliminary dimensionality reduction of $\vec{Q}$ by means of the Principal Component Analysis \citep[PCA,][]{doi:10.1080/14786440109462720}, obtaining pre-processed `morphology' vectors, $\vec{\theta}_{\rm PCA}$. PCA is a method that orthogonally projects the data into a coordinate system in which successive vector components explain a smaller and smaller part of the data variance. The target number of its dimensions returned by PCA for each $\vec{Q}$ was set to 10. This value was chosen experimentally by examining the percentage of the total variance of the data set explained by successive PCA components. For $\mathcal{L}_1(t)$, the first ten PCA components explained a total of 99.8\% of variance (first component -- 79\% and second component -- 19\%). For $\mathcal{L}_2(t)$, the corresponding value was 98.5\% (in this case, the first PCA component explained 60\% of the total variance, while the second explained 22\%).

We then performed the final 2D embedding by applying UMAP on the collection of $\vec{\theta}_{\rm PCA}$ vectors. UMAP is a multipurpose non-linear dimensionality reduction technique that constructs a low-dimensional projection that preserves as accurately as possible the topological structure of the input data. For instance, in this case, a pair of embeddings of resonance curves with similar properties (in the sense of their summary statistics described above) are expected to lie in mutual vicinity on the 2D UMAP plane. Let us denote the UMAP results as $\vec{\theta}_{\rm UMAP}$. The manifold spanned by $\vec{\theta}_{\rm UMAP}$ (Sect.~\ref{sect:results-Investigation of the morphology of resonance curves using UMAP}) allowed us to effectively examine the differences in the morphology of the resonance curves and their dependence on the initial parameters of the simulated EEVs.

Unlike PCA, UMAP is a complex method, with many free parameters that need to be adjusted with care, as the resulting embedding may depend heavily on their choice. Appendix~\ref{appendix:umap} provides all the `technical' details of this process, including the values of the most important UMAP parameters adopted in this study.

%-------------------------------------------------------------------
\section{Results}\label{sect:results}

\subsection{General properties of synthesised EEVs}
Before going into a detailed analysis of the resonance curves, we briefly characterise the general properties of the models we have synthesised using the \texttt{MESAbinary} and \texttt{GYRE} codes.

\subsubsection{Evolutionary tracks in HRD}
\begin{figure*}
   \centering
   \includegraphics[width=\hsize]{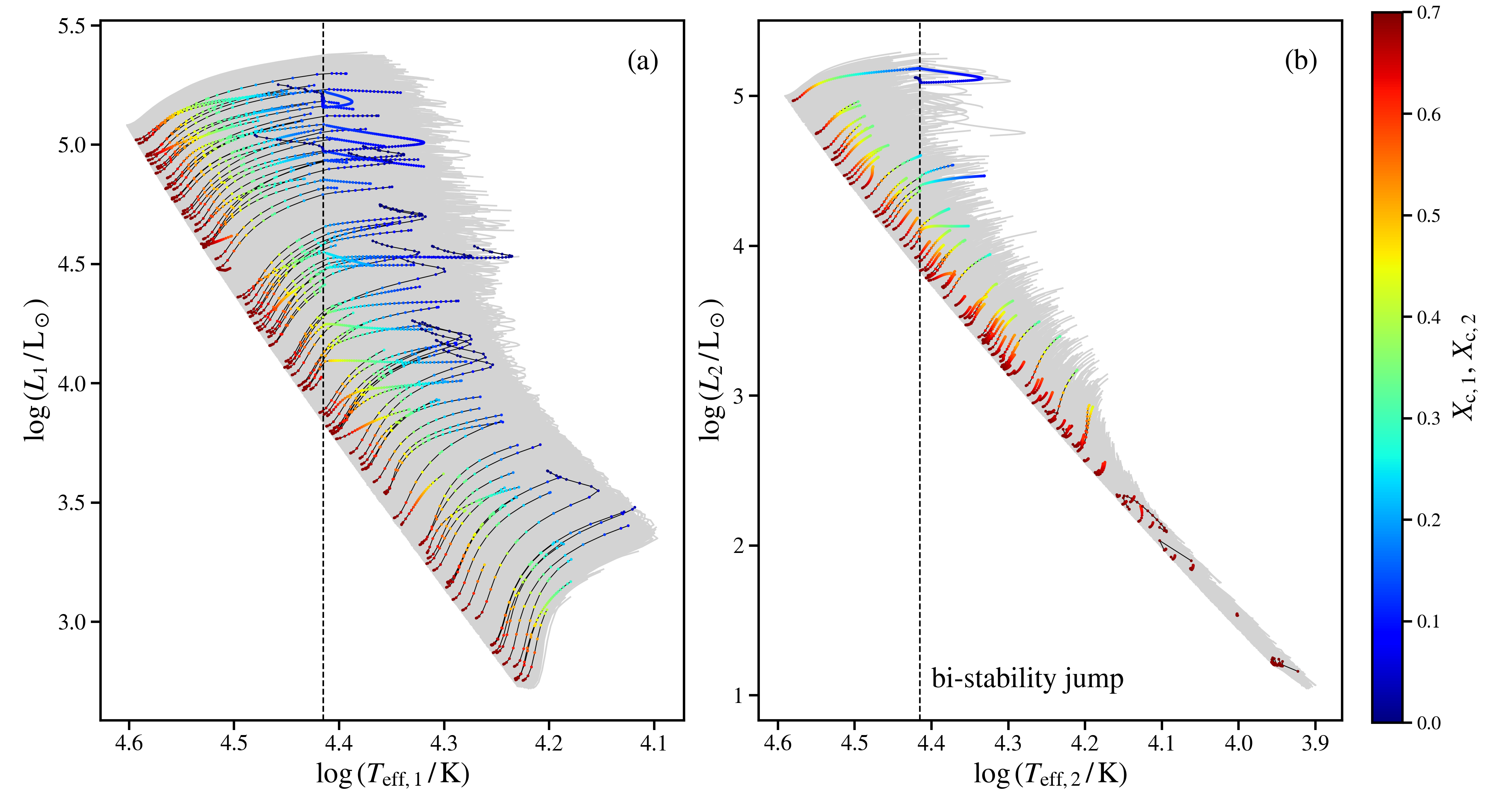}
   \caption{(a) HRD with the evolutionary tracks of primary components. The grey area corresponds to the region occupied by the full set of 20,000 tracks, while a subsample of 100 randomly-selected tracks is indicated with coloured points connected by black solid lines. Each point represents one saved \texttt{MESA} model. The colour coding reflects the central hydrogen abundance. The effective temperature of the bi-stability jump ($T_{\rm eff}\approx$\,26,000\,K) is marked with the vertical dashed line. The abrupt change in the behaviour of some evolutionary tracks after crossing the bi-stability jump region is due to a significant change in the wind mass-loss rate; (b) The same as panel (a), but for a set of secondary components. We note the difference in the ranges of the two axes in panels (a) and (b). More details are discussed in the main text.}
   \label{fig:hrd}
\end{figure*}

Figure~\ref{fig:hrd} shows a pair of HRDs with a compilation of all 20,000 evolutionary tracks that we obtained in our simulations for the primary (Fig.~\ref{fig:hrd}a) and secondary (Fig.~\ref{fig:hrd}b) components. Although it is impossible to clearly present thousands of evolutionary tracks on a single HRD, we have highlighted and colour-coded a small fraction of them in order to describe some of their features.

First of all, only a fraction of the primaries reached TAMS when the central mass abundance of hydrogen dropped below $10^{-4}$ (according to the first of our termination conditions, Sect.~\ref{sect:methods-termination}). Many evolutionary tracks were interrupted at MS due to the fulfilment of one of the other termination conditions. Secondly, a number of tracks clearly change their character after crossing the line corresponding to the bi-stability jump (around $T_{\rm eff}=$\,26,000\,K). This is due to the associated sharp increase in the wind mass-loss rate, as it tries to keep the stellar luminosity constant. In some circumstances, the mass-loss rate is so high that the star loses a significant part of its envelope\footnote{We recall that these high mass-loss rates are not exclusively derived from the description of \cite{2001A&A...369..574V}. Rotational and tidal amplification mechanisms can significantly intensify stellar winds in our simulations. These phenomena are particularly well-pronounced when the component is close to TAMS, i.e., its radius approaches the Roche-lobe radius.}. This effect `pushes' the star back to the high effective temperature region and is particularly pronounced for the most massive stars in our sample (cf.~Fig.~\ref{fig:hrd}a, evolutionary tracks that `turn around' and cross the bi-stability jump for a second time).

\subsubsection{EEV groups in terms of the termination condition}
Only four of the seven\footnote{In Sect.~\ref{sect:methods-termination} we give four types of termination conditions, but three of them apply independently to both the primary and secondary components.} termination conditions actually occurred in our simulations. The majority of our EEVs ($\sim$67.1\,\%) ended up as MS RLOF systems in which the primary component filled its Roche lobe during the periastron passage. The next most numerous group ($\sim$22.6\,\%) were systems in which the primary component successfully reached TAMS ($X_{\rm c}\leqslant 10^{-4}$). About 10.2\,\% of the binaries managed to circularise their orbits before any other termination condition was met. The last group contains only about 0.04\,\% of the total sample. This is the group where the primary's rotation velocity exceeded the maximum allowed angular velocity ($\Omega\,/\,\Omega_{\rm crit}\geqslant 0.75$).

\begin{figure}
\centering
\includegraphics[width=0.95\hsize]{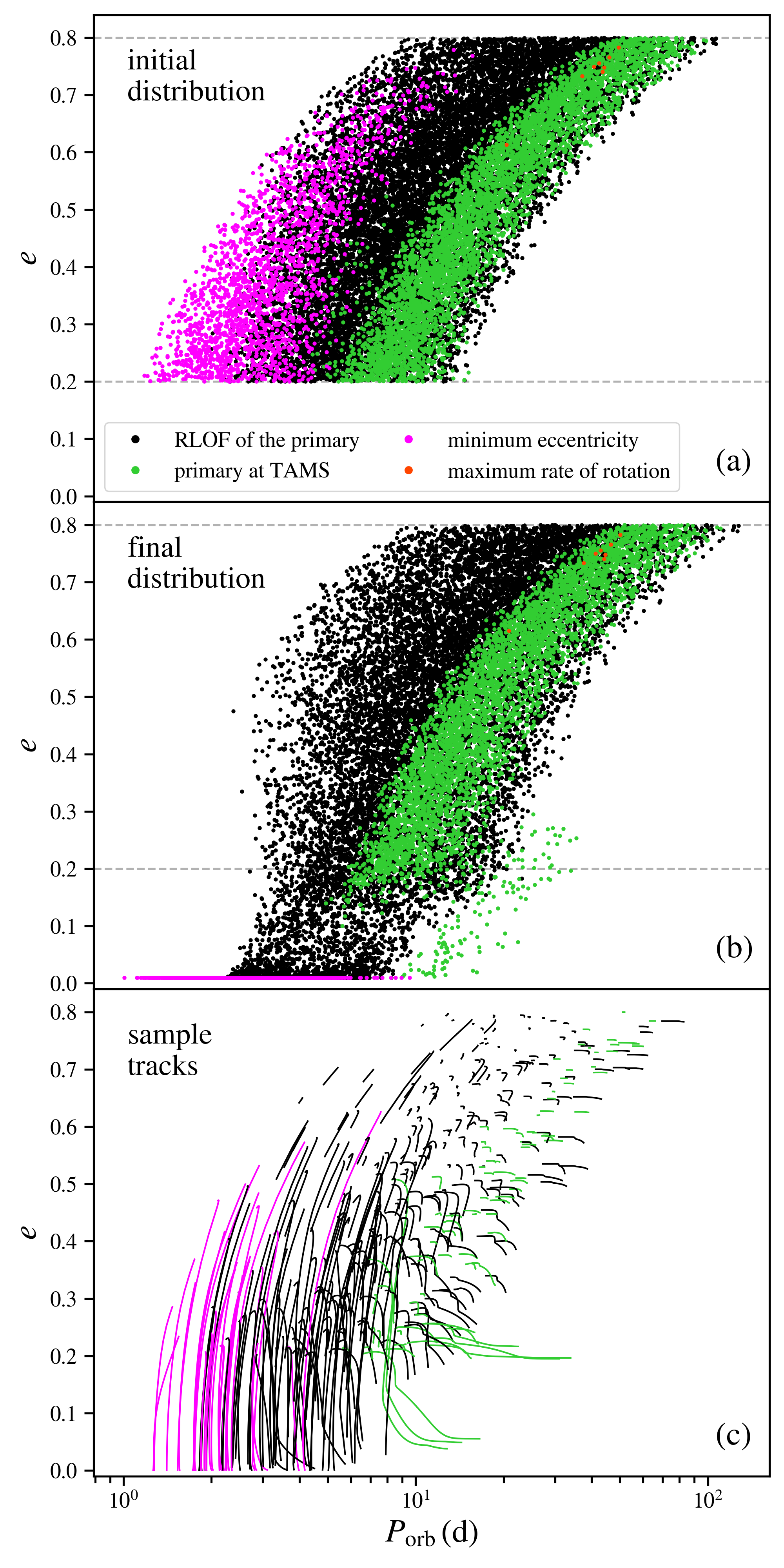}
\caption{Orbital period-eccentricity distributions of 20,000 modelled EEVs; (a) Initial distribution of $e$ as a function of $P_{\rm orb}$. Colour-coding corresponds to the termination conditions described in Sect~\ref{sect:methods-termination}, i.e. the RLOF of the primary component during periastron passages before reaching TAMS (black), exhaustion of hydrogen in the primary's core (primary at TAMS, green), almost complete circularization of the orbit ($e=0.01$, magenta), and the maximum allowed rotation rate of the primary component ($\Omega\,/\,\Omega_{\rm crit}=0.75$, orange). A pair of dashed horizontal lines mark the boundary values of the initial eccentricity, $e=0.8$ and $e=0.2$; (b) Same as in panel (a), but for the final state of each modelled binary system; (c) Random selection of 400 orbital evolution tracks with the same colour-coding as in panels (a) and (b).}
\label{fig:P-e-final-plane}
\end{figure}

Figure~\ref{fig:P-e-final-plane} presents these four groups of EEVs on the $P_{\rm orb}$-$e$ plane and allows a comparison of the initial (Fig.~\ref{fig:P-e-final-plane}a) and final (Fig.~\ref{fig:P-e-final-plane}b) states of the aforementioned distribution. As expected, the EEVs with the shortest orbital periods and high eccentricities tended to circularise their orbits before leaving the MS. Their trajectories in the $P_{\rm orb}$-$e$ diagram (Fig.~\ref{fig:P-e-final-plane}c) follow smooth, almost vertical lines due to the strong tidal damping of eccentricity. On the other hand, the integration of the evolution of systems with large distances between components at periastron ($\widetilde{r}_{\rm peri}\gtrsim 3.5$) has been terminated mainly due to the exhaustion of hydrogen in the primary's core. Although the majority of EEVs belonging to this group do not significantly change their orbital parameters during evolution, there is a subgroup of them that behaves quite differently. It can be recognised as the distinct `cloud' of green dots in Fig.~\ref{fig:P-e-final-plane}b, represented by the mainly horizontal green tracks in Fig.~\ref{fig:P-e-final-plane}c. These are systems that were characterised by very strong stellar winds at the end of the MS phase and have lost much of their envelopes, so that their orbital period has increased significantly (Kepler's third law).

The most numerous group of EEVs, in which the primary component has filled its Roche lobe in the MS phase, forms a kind of `bridge' between the two previously mentioned groups and merges with them. The shapes of the corresponding trajectories on the $P_{\rm orb}$-$e$ plane may vary from system to system, depending on the interplay between tidal forces and the intensity of stellar winds, so no single `type' of track can be assigned to them. However, many of them resemble the inverted Greek letter `$\Gamma$' -- initially, the system drifts horizontally (towards the longer orbital period) as a result of the mass loss and/or spin-orbit coupling, and then undergoes more or less rapid circularisation (moves vertically downwards) under the influence of the intense tides, which come to the fore when the primary component almost fills its Roche lobe.

Only 8 out of 20,000 EEVs underwent efficient spin up of both components due to the pseudo-synchronisation \citep[when the rotation period of the star `matches' the rate of orbital motion at periastron, so that there is no net torque over an orbital cycle, e.g.,][]{1981A&A....99..126H}. These few systems are located in the upper right corner of Figs.~\ref{fig:P-e-final-plane}a and b. Their orbits were initially highly eccentric yet relatively widely-separated at periastron ($\widetilde{r}_{\rm peri}\approx4.5$\,--\,5.0). Thus, in combination with the lower masses of the primary components ($M_1\approx 5\,M_\odot$), there was no effective tidal dissipation. However, the envelopes of these stars tended to rotate pseudo-synchronously with the orbit (due to the relatively long nuclear time scale of the evolution of a 5\,$M_\odot$, the primaries had enough time to do so). Consequently, this led to a very fast rotation of the primary component, exceeding the threshold of $\Omega/\Omega_{\rm crit}=0.75$.

\subsubsection{Internal structure and asteroseismic properties}
\begin{figure*}
   \centering
   \includegraphics[width=0.95\hsize]{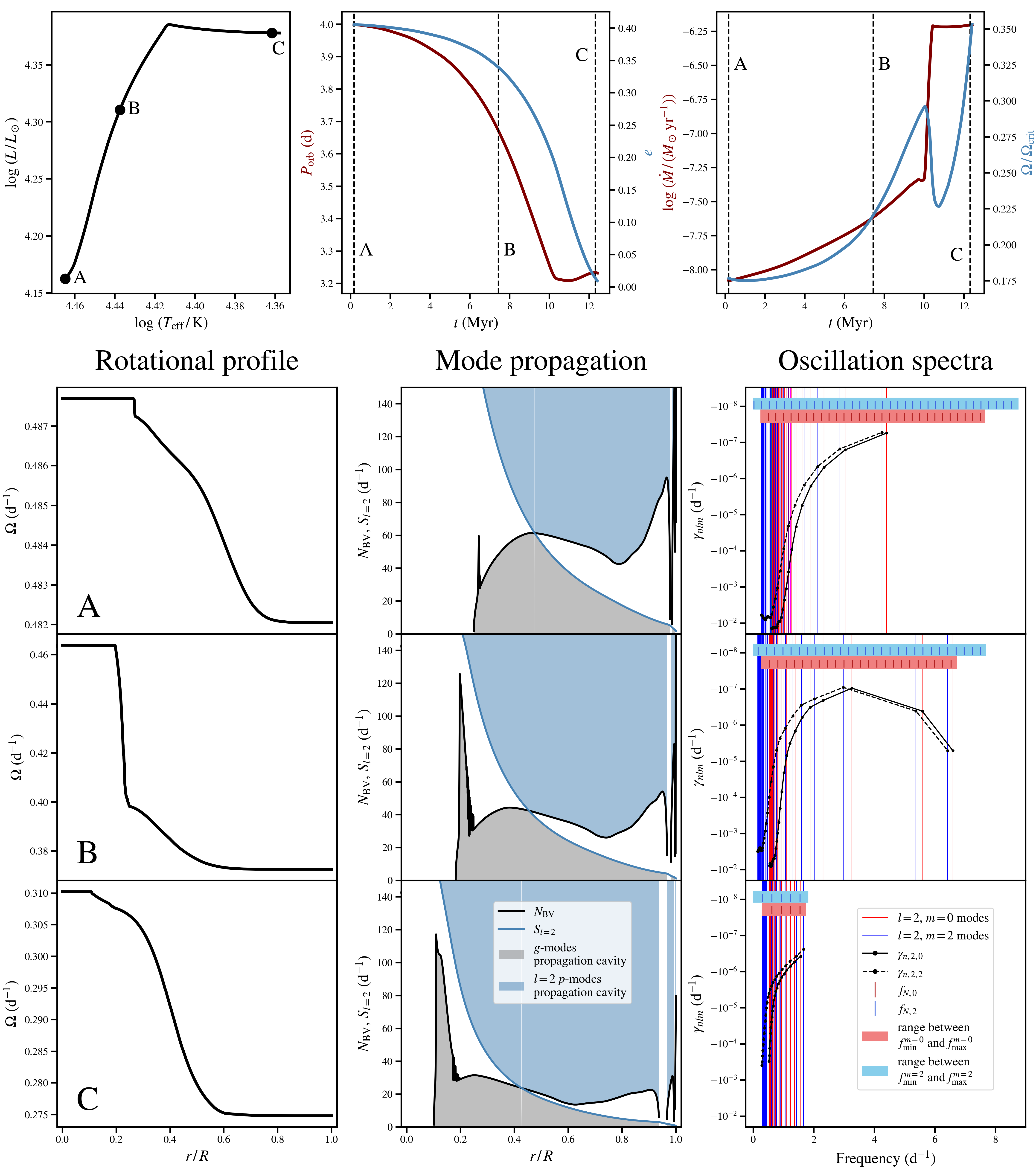}
   \caption{Summary plot of the properties of the primary component of one of the arbitrarily selected binary systems from our simulations. The approximate initial parameters of this particular system were as follows: $M_1\approx13.6\,M_\odot$, $M_2\approx3.6\,M_\odot$, $e\approx0.4$, and $\widetilde{r}_{\rm peri}\approx2.3$. The integration of the system was terminated because of its circularisation. The top row of panels shows, from left to right, evolutionary track in the HRD, the evolution of the orbital period and eccentricity, and the temporal changes of the wind mass-loss rate and surface rotation velocity. The vertical dashed lines in the latter two diagrams correspond to epochs A, B, C in the HRD. The lower part of the figure shows the internal rotation profile (left column), the mode-propagation diagram (middle column) and the synthetic oscillation spectrum (right column) for epochs A, B, C (shown in consecutive rows labelled with these letters). The rotation frequency inside the primary is drawn as a function of fractional radius, $r\,/\,R$. The range of rotation frequency is different in the three panels. The mode-propagation diagram shows the dependence of the Brunt-V\"ais\"al\"a frequency (black line) and Lamb frequency for $l=2$ modes (blue line) on the fractional radius. The grey and blue shaded areas correspond to the propagation cavities of the $g$ and $l=2$ $p$ modes, respectively. The synthetic oscillation spectrum diagrams contain several different pieces of information. The light blue and light red horizontal bars delineate the range of frequencies allowed by the $F_{Nm}$ values. In the background of each, the blue and red short vertical lines indicate tidal-forcing frequencies lying within these ranges. The synthetic oscillation spectra calculated by \texttt{GYRE} are marked with red ($\sigma_{n,2,0}$) and blue ($\sigma_{n,2,+2}$) long vertical lines. Their corresponding linear damping rates are plotted as solid ($\gamma_{n,2,0}$) and dashed ($\gamma_{n,2,2}$) black lines. The frequency scale on the abscissa axis refers to the rest frame co-rotating with the primary's core.}
   \label{fig:all-in-one}
\end{figure*}

The shape of the resonance curve depends not only on the global properties of the components and the orbit, but also on the internal structure of the stars, which directly affects seismic properties (i.e. the spectrum of eigenmodes). Therefore, within the limited volume of this paper, we would like to show at least one representative example of the evolution of the internal properties of the primary component for an arbitrarily-chosen EEV. Figure~\ref{fig:all-in-one} shows the evolution of a primary with mass $M_1\approx13.6\,M_\odot$ in a system with an initial eccentricity $e\approx0.4$ and an initial orbital period $P_{\rm orb}\approx4.0\,$d. In our simulations, this particular system finished its evolution due to the circularisation of its orbit after about 12\,Myrs. The HRD in Fig.~\ref{fig:all-in-one} reveals the `non-standard' evolutionary track of the primary due to the sharp change in the mass-loss rate after crossing the bi-stability jump (right panel in the top row of Fig.~\ref{fig:all-in-one}). The same panel also shows how the primary's surface rotation rate varies over time -- as the mass-loss rate increases, it loses a lot of spin angular momentum and slows down its rotation. The eccentricity and orbital period monotonically decrease with time (middle panel in the top row of Fig.~\ref{fig:all-in-one}), except for a short episode of increase in $P_{\rm orb}$ caused by the irreversible loss of a large part of the envelope. We have selected three epochs in the evolutionary history of this EEV (labelled A, B, C on the HRD), for which we have presented the appearance of the rotational profiles, mode propagation diagrams and oscillation spectra of the primary component in the bottom part of Fig.~\ref{fig:all-in-one}. Epoch A corresponds to the phase of evolution just after leaving the ZAMS, epoch B is characterised by $X_{\rm c,1}\approx 0.45$, and finally, epoch C marks the situation just before the complete circularisation of the EEV. Let us briefly describe the changes occurring in each of the three types of diagram below. 

The internal rotation profile of the primary is almost constant for epoch A, but by then a division between a faster-rotating core and a slower-rotating envelope begins to emerge. The aforementioned division becomes particularly apparent in epoch B, when the core has developed a rotation rate approximately 1.25 times that of the surface layers. As can be seen, the contracting core rotates as a rigid body throughout the MS lifetime due to efficient angular momentum transport supported by convection. The outer part of the envelope also rotates almost rigidly, but this time it is due to large-scale Eddington-Sweet meridional flows. The angular velocity gradient in the primary starts to gradually decrease as the star reaches epoch C. Various mixing processes in the chemically-modified layer left by the core lead to the diffusion of angular momentum from the core to the envelope. Moreover, the rotational profile inside the star becomes a smooth function of the radius (rather than a step-like function as for epoch B).

The majority of TEOs in our simulations belong to the $g$-mode family of oscillations, so it is very important to control the behaviour of the Brunt-V\"ais\"al\"a buoyancy frequency, $N_{\rm BV}$, in our models. Together with Lamb frequency for $l=2$ modes, $S_{l=2}$, they carry information about $g$ and $p$ mode cavities and their evanescence regions \citep[e.g.,][their Sect.~3.4]{2010aste.book.....A}. The evolution of $N_{\rm BV}$ and $S_{l=2}$ is presented in the middle column of Fig.~\ref{fig:all-in-one}. The blue and grey shaded regions denote the position of the $l=2$ $p$-mode and $g$-mode propagation cavities, respectively. The white areas that lie between the Brunt-V\"ais\"al\"a and Lamb frequencies correspond to the evanescence regions. During evolution, the receding convective core builds up a large $g$-mode trapping cavity, which is very important for their frequency spectrum. Additionally, the behaviour of the $N_{\rm BV}$ just below the stellar photosphere reveals a pair of thin subsurface convection zones, expected for this type of star \citep[e.g.,][]{2022ApJS..262...19J}. Comparing the mode propagation diagrams for epochs A and C, it can be seen that also the $p$ modes can penetrate deeper and deeper into the primary as it gradually depletes the hydrogen in its core.

The right column in Fig.~\ref{fig:all-in-one} contains most of the information that is directly used to obtain the resonance curve. The horizontal bars at the top of each panel correspond to the frequency range in which \texttt{GYRE} looked for potential TEOs (according to the criteria adopted in Sect.~\ref{sect:asteroseismic-calculations}). We recall that that their width depends on the $N_{\rm max}^m(e)$ functions, so as the system evolves towards lower eccentricities, these bars are shorter and shorter (i.e. fewer harmonics of the orbital frequency can effectively drive TEOs). With the thick, short vertical lines we mark the location of the tidal forcing frequencies. As can be seen, the separation between successive values of $f_{Nm}$ becomes larger with passing time due to the increase in $f_{\rm orb}$. The eigenfrequencies found by \texttt{GYRE} are marked with the long thin vertical lines, while the linear damping rates of these modes are shown as black solid and dotted lines. The presented set of three synthetic oscillation spectra reveals a typical structure for $g$ modes with their asymptotic behaviour for large radial orders (which correspond to lower frequencies). It may appear that the dense `forests' of eigenfrequencies end too early relative to the left limits of horizontal bars. However, this is not a mistake, but a direct consequence of the maximum $|n|$ we allowed in the calculations -- modes with lower frequencies would have larger radial orders than thirty. During the evolution of the EEV, both the forcing frequencies and the oscillation spectrum shift, so the intersection of these two vertical line patterns is virtually inevitable in most cases. Each of these intersections is the source of a single resonance that can give rise to a noticeable TEO at the level of the photosphere.

\subsection{The `visual inspection' of resonance curves}\label{sect:results-eye-inspection}
The resonance curves are characterised by a striking diversity in terms of morphology, which is already partly evident in Fig.~\ref{fig:resonance-curves-sample}. The four examples of $\mathcal{L}_1(t)$ and $\mathcal{L}_2(t)$ shown in this figure show that the components of the EEVs can experience, firstly, a very different number of resonances and, secondly, their distribution in time can take various forms. The heights of the maxima of the resonance curves are mainly dictated by the $\gamma_{nlm}$ of the mode to which the smallest difference corresponds, $(\sigma_{nlm}-f_{Nm})$. Statistically speaking, modes with larger $|n|$ are more strongly non-adiabatic (have larger damping rates), hence the maxima they induce in the resonance curves are lower (cf.~Eq.~(\ref{eq:L_N})). Another factor determines the extent of the resonant maximum in time. It is determined by the relative `velocity' with which the eigenfrequency spectrum crosses the $f_{Nm}$ spectrum. By `velocity' here we mean the rate of change of these two independent frequency spectra.

It should be emphasised that there are also numerous cases in which $\mathcal{L}(t)$ drops sharply to zero at some point (cf. $\mathcal{L}_2(t)$ curve in the bottom panel of Fig.~\ref{fig:resonance-curves-sample}) or resonances do not occur at all (see Sect.~\ref{sect:results-total-number-of-resonances} and Fig.~\ref{fig:total-number-of-resonances}). Such a situation can occur, for example, when the oscillation spectrum lies completely outside the frequency range allowed by the $F_{Nm}$ coefficients or the nuclear timescale of the secondary is much longer than the same time scale for the primary. Under such circumstances, the secondary component will remain close to the ZAMS until the termination condition is met. Thus, it will not significantly change its internal structure and oscillation spectrum. This in turn means that the oscillation spectrum will not move relative to the tidal forcing frequencies, effectively reducing the number of possible resonance events.

\subsubsection{`Long' resonances}\label{sect:results-long duration resonances}
\begin{figure}
\centering
\includegraphics[width=\hsize]{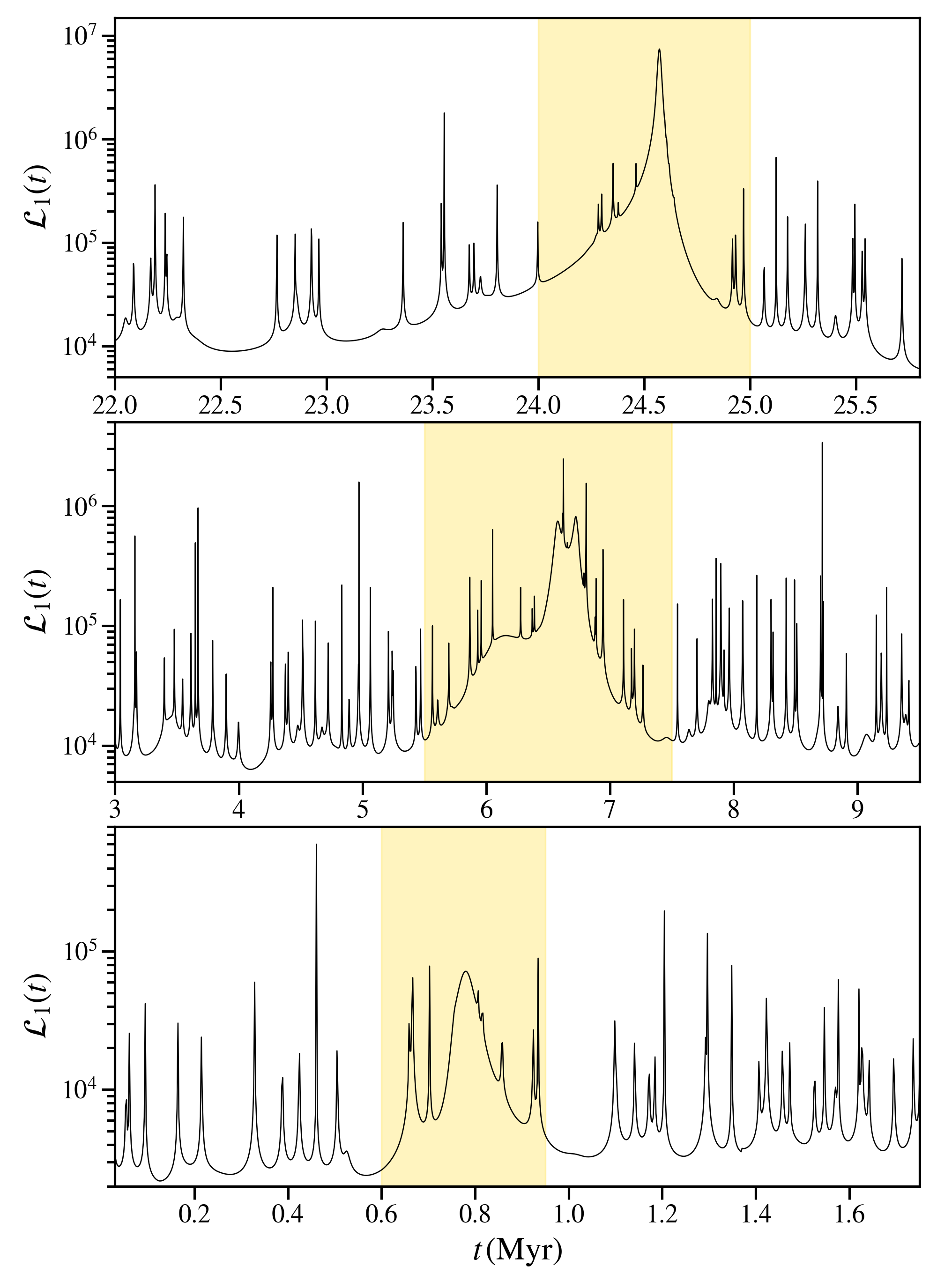}
\caption{Example resonance curves of the primary component for three different EEVs from our simulations that exhibit long resonances (highlighted by the shaded areas in each panel). We note a substantial difference in the width of typical and long resonances. These long resonances are good candidate for excitation of high-amplitude or resonantly-locked TEOs.}
\label{fig:long-resonances}
\end{figure}

Our sample of resonance curves includes a particular group of $\mathcal{L}(t)$ curves that exhibit exceptionally long duration resonances compared to typical ones (we will refer to them as `long resonances'). Figure~\ref{fig:long-resonances} presents parts of three representative resonance curves belonging to this group. The shaded regions in the figure mark the position of the long resonances. As can be clearly seen, the typical resonance usually lasts for about $10^3$\,--\,$10^4$\,years, which is approximately 100 times shorter than the duration of a long resonance (of the order of $10^5$\,--\,$10^6$\,years). They originate from the intersection of one of the $f_{Nm}$ frequencies with the $\sigma_{nlm}$ frequency at a very small angle, in terms of their temporal evolution. As a result, they remain for a relatively long time in very close vicinity, leading to a broad resonance overlapping with narrower ones (originating from other intersections of the $f_{Nm}$ and $\sigma_{nlm}$ frequency spectra; cf. especially the middle panel in Fig.~\ref{fig:long-resonances}). The long resonances are interesting for at least two reasons. First of all, they are natural candidates for resonantly-locked TEOs. However, based on our simulations, it is difficult to say whether an extended resonance would persist when the back-reaction of a TEO on the orbit is taken into account. Secondly, if the energy exchange between the eigenmode and the orbit that corresponds to a long resonance is not efficient (i.e. there is a small chance that a long resonance will be lost), such a resonance should lead to a high-amplitude TEO without the need for resonance locking. This is simply because it has enough time to reach its saturation level due to the non-linear effects. However, we did not find any significant correlations between the occurrence of a long resonance in $\mathcal{L}(t)$ and the initial parameters of our EEVs.

\subsection{Total number of resonances and the average rate of their occurrence}\label{sect:results-total-number-of-resonances}
\begin{figure*}
   \sidecaption
   \includegraphics[width=12cm]{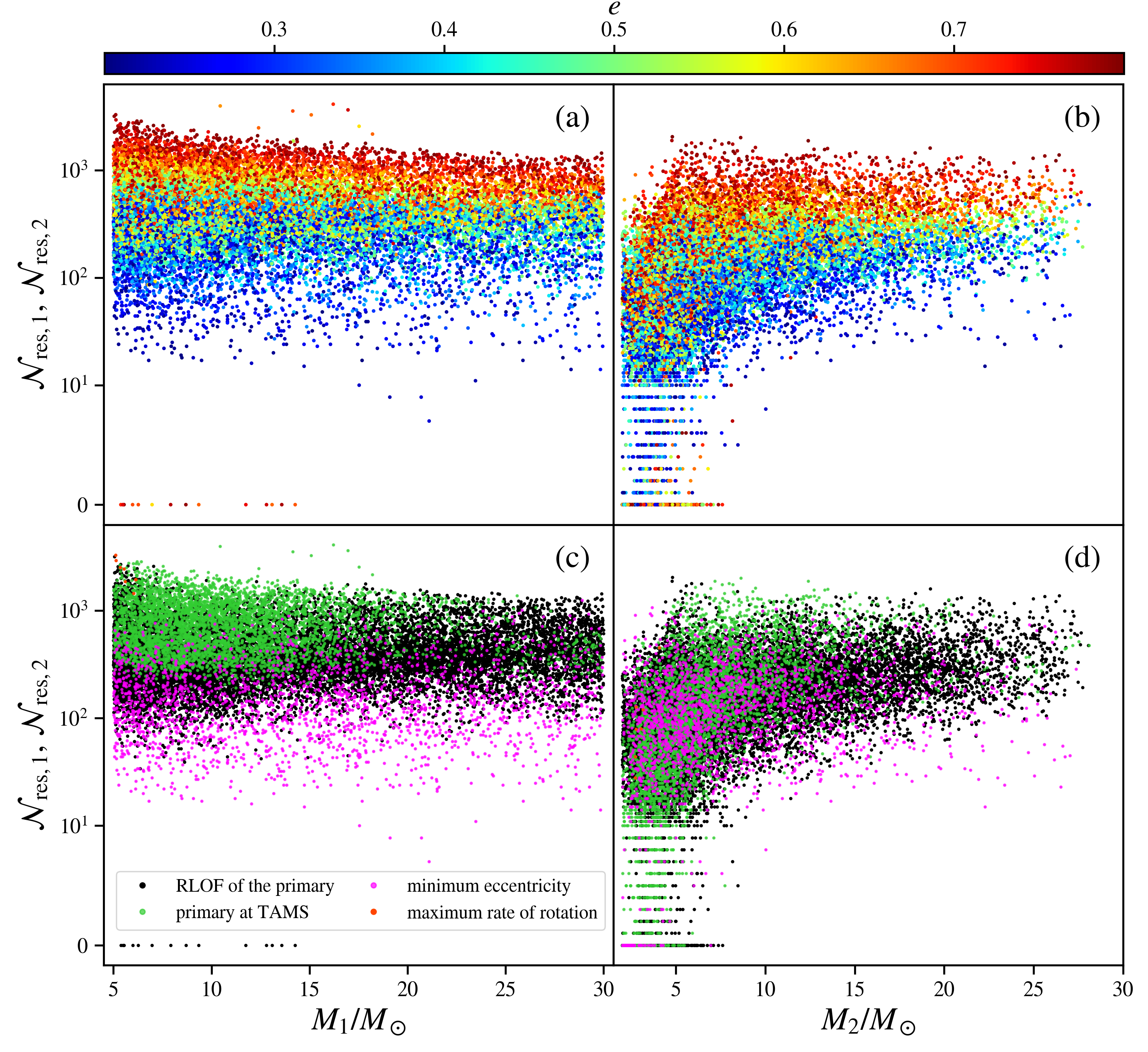}
   \caption{Total number of resonances in the primary (a, c) and secondary (b, d) components that we detected in our simulations as a function of the initial masses of the components. The initial eccentricity of the EEV is colour-coded in panels (a) and (b), while the corresponding scale is shown at the top of the figure. Panels (c) and (d) are analogous to their counterparts in the top row, but the colours used reflect the termination criterion (same as in Fig.~\ref{fig:P-e-final-plane}). The ordinate scale is logarithmically-scaled for $\mathcal{N}_{\rm res}>10$. Below this value, a linear scale was applied in order to present components without resonances, i.e.~$\mathcal{N}_{\rm res,1}$ or $\mathcal{N}_{\rm res,2}$ equal to zero.}
   \label{fig:total-number-of-resonances}
\end{figure*}

The first feature of the morphology of the resonance curves that we investigated is the total number of resonances that occurred in the primary and secondary components, $\mathcal{N}_{\rm res,1}$ and $\mathcal{N}_{\rm res,2}$. However, we did not calculate these statistics directly from $\mathcal{L}_1(t)$ and $\mathcal{L}_2(t)$, because some of the apparent maxima may actually be a blend of more than one resonance event. This is especially true when the involved $\gamma_{nlm}$ differ by orders of magnitude. Then one of the resonances is characterised by a notably smaller maximum, which seems to `hide' in the dominant one. Instead, we used a different approach that did not underestimate the actual number of resonances. When post-processing the generated models, we simply counted each intersection of the $\sigma_{n,2,0}$ and $\sigma_{n,2,+2}$ frequency spectra with their counterparts $f_{N,0}$ and $f_{N,2}$, respectively. The results of such an analysis are depicted in Fig.~\ref{fig:total-number-of-resonances}.

The most important thing about Fig.~\ref{fig:total-number-of-resonances} is that it shows the absolute number of resonances. EEVs can experience hundreds or even thousands of resonances during their evolution on MS. It would therefore be wrong to claim that these phenomena are rare in massive and intermediate-mass EEVs, although in general, resonances are quite short-lived compared to the nuclear time scale. The total number of resonances experienced by the primary component (Fig.~\ref{fig:total-number-of-resonances}a) shows a correlation with both its initial mass and the initial eccentricity of the system. The mildly decreasing trend of $\mathcal{N}_{\rm res,1}$ towards higher $M_1$ originates from the fact that the mean lifetime of the star on MS shortens with increasing mass. On the other hand, the wide range of $\mathcal{N}_{\rm res}$ is mainly due to differences in initial eccentricity. The closer the system is to a circular geometry at the beginning of evolution, the statistically lower the value of $\mathcal{N}_{\rm res}$, which is self-explanatory and also applies to the secondary component (Fig.~\ref{fig:total-number-of-resonances}b). EEVs that have managed to circularise their orbits in the MS phase (magenta dots in Fig.~\ref{fig:total-number-of-resonances}c) have on average lower initial eccentricities and thus fewer resonances. The opposite behaviour is exhibited by EEVs in which the primary component has had a chance to reach TAMS (green dots in Fig.~\ref{fig:total-number-of-resonances}c). The secondary components experience a slightly fewer resonances compared to the primaries (Fig.~\ref{fig:total-number-of-resonances}b) and there is no clear division of the $\mathcal{N}_{\rm res,2}$ distribution with respect to the termination criterion (Fig.~\ref{fig:total-number-of-resonances}d). The noticeably smaller number of resonances for secondary components with masses $M_2<5\,M_\odot$ comes from the conditions of our simulations, i.e. secondaries with these masses occur in systems with decreasing mass ratios\footnote{We recall that the minimum mass of the primary component considered in our study was equal to $5\,M_\odot$.}. Hence, the large difference in nuclear time scales between the components means that the secondary component does not significantly change its eigenfrequency spectrum, resulting in a smaller number of resonances.

As we already mentioned in Sect.~\ref{sect:results-eye-inspection}, some of the $\mathcal{L}_1(t)$ and $\mathcal{L}_2(t)$ curves do not reveal any resonances, which is why they lie in Fig.~\ref{fig:total-number-of-resonances} on the horizontal line $\mathcal{N}_{\rm res}=0$. This behaviour occurred for only 0.07\% of our primaries. They are all EEVs with highly eccentric orbits that quickly filled their Roche lobes at periastron. There was much more such behaviour for secondaries, about 7\%, mainly for the intermediate-mass companions of the much more massive primaries.

\begin{figure*}
   \sidecaption
   \includegraphics[width=12cm]{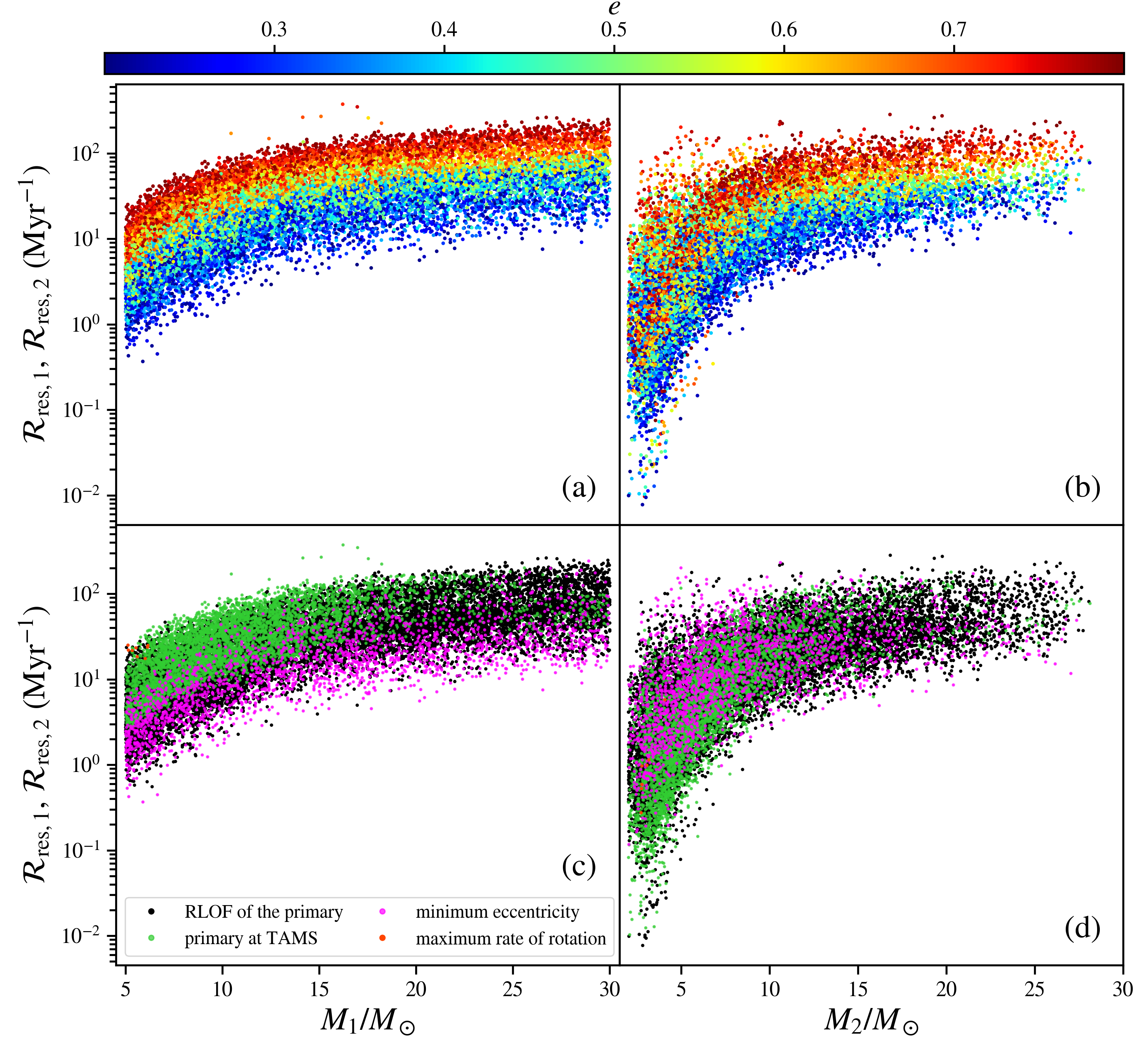}
   \caption{Summary plots analogous to Fig.~\ref{fig:total-number-of-resonances}, but showing the average rate of resonances occurring in the simulated EEVs (average number of resonances per Myr). Components that did not exhibit any resonances during the simulation have been omitted here as their $\mathcal{R}_{\rm res}$ value would simply be zero. The colour-coding is the same as in Fig.~\ref{fig:total-number-of-resonances}.}
   \label{fig:number-of-resonances-per-myr}
\end{figure*}

From an observational point of view, even more important than the total number of resonances is the rate at which they occur. Knowing this rate, for a given population of MS EEVs, we can approximately say where we have a statistically higher chance of observing TEOs. After all, our observations only correspond to one particular moment in time, not the entire evolution. Knowing the values of $\mathcal{N}_{\rm res}$ for each component and the age of each system at termination, $T_{\rm max}$, we can calculate the average rate of resonances as $\mathcal{R}_{\rm res}\equiv \mathcal{N}_{\rm res}/T_{\rm max}$. We show the distribution of $\mathcal{R}_{\rm res,1}$ and $\mathcal{R}_{\rm res,2}$ in Fig.~\ref{fig:number-of-resonances-per-myr}. It is very difficult to predict what the dependence of $\mathcal{R}_{\rm res}$ on the mass of the component will look like, as it is the result of a complex interplay between many related factors. On the one hand, it can be said that massive stars should have a smaller $\mathcal{R}_{\rm res}$ because their lifetimes are shorter and they fill their Roche lobes relatively easily (in the considered range of orbital parameters). On the other hand, however, massive stars quickly change their internal structure (i.e.~asteroseismic properties), so that their eigenfrequency spectra evolve rapidly, increasing the likelihood of interaction with the structure of the tidal forcing frequencies. The question is, which of these processes prevails? As can be seen in Fig.~\ref{fig:number-of-resonances-per-myr}, it is the more massive stars that are more likely to undergo resonances. Both primary and secondary components with masses around $30\,M_\odot$ have on average an order of magnitude higher $\mathcal{R}_{\rm res}$ ($\sim$$10^2$\,Myr$^{-1}$) than components with masses around $5\,M_\odot$ ($\sim$$10^1$\,Myr$^{-1}$, Fig.~\ref{fig:number-of-resonances-per-myr}a and b). Moreover, the dependence of the distributions shown in Fig.~\ref{fig:number-of-resonances-per-myr} on the initial eccentricity and termination condition is inherited from Fig.~\ref{fig:total-number-of-resonances}.

At this point, we can venture the conclusion that in the case of MS EEVs, TEOs should be observed mostly in the upper part of the MS (among early B- and O-type dwarfs), which still requires observational verification on a large sample of massive EEVs. Although we cannot extrapolate the obtained distributions of $\mathcal{R}_{\rm res}$ towards lower masses, these stars have an increasingly extended convective envelope, which in turn should effectively limit the photometric detection of $g$-mode TEOs. On the contrary, the envelopes of massive stars are radiative, which should not prevent $g$-mode TEOs from propagating up to the vicinity of the photosphere. Thus, they can be more easily detected by analysing the light curves, especially in the era of high-quality space-borne photometry.
 
\subsection{Distribution of resonances over time}\label{sect:results-Distribution of resonances over time}
\begin{figure}
\centering
\includegraphics[width=1\hsize]{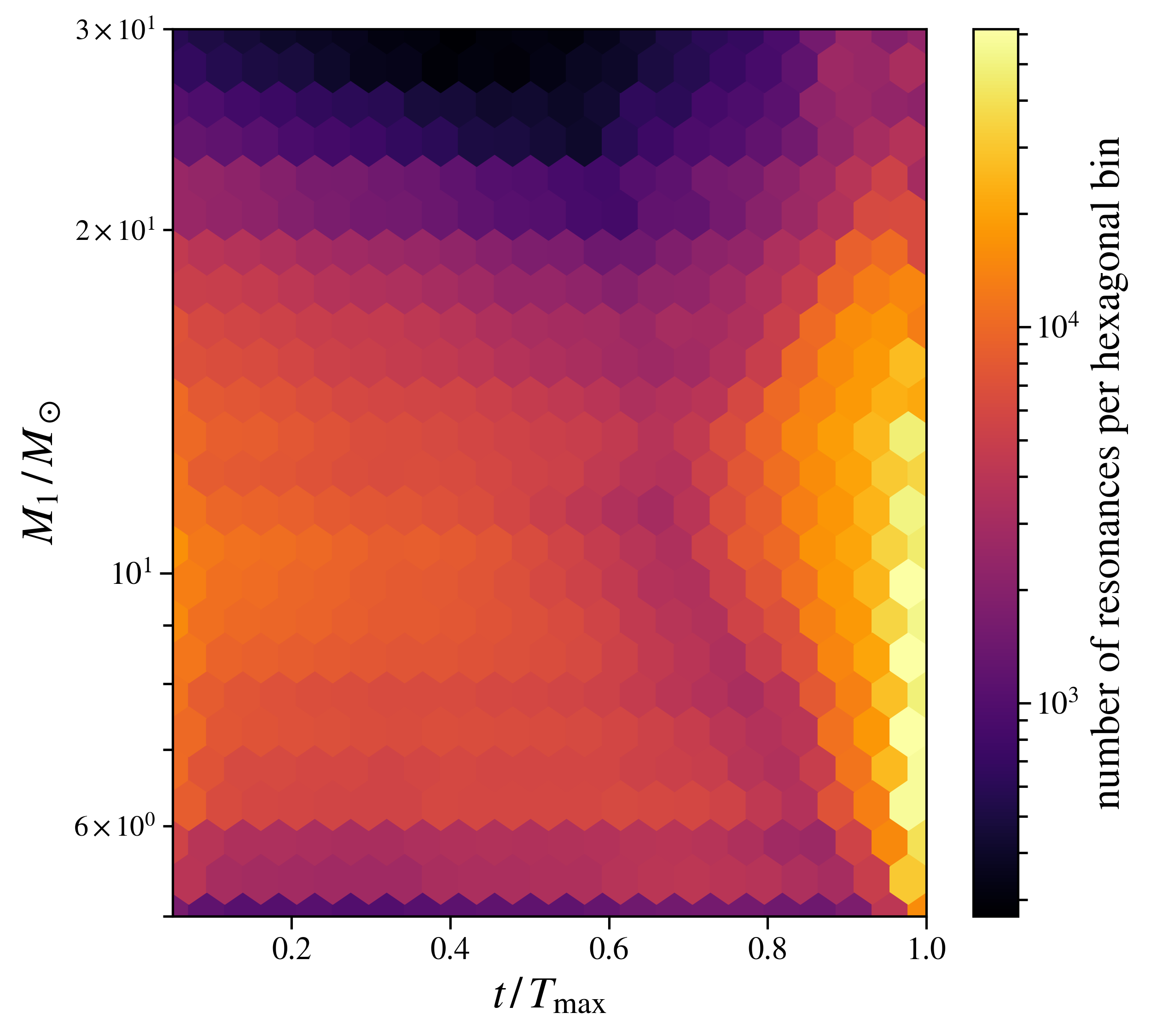}
\caption{Time distribution of the resonances of the primary component that reached TAMS. The abscissa axis corresponds to the normalised time and the ordinate shows the initial mass of the primary component. In addition, the ordinate is logarithmically scaled, so the set of resonance curves is almost uniformly distributed in the vertical direction. The total number of resonances contained in one hexagonal bin is colour-coded according to the scale on the right.}
\label{fig:time-series-of-resonances-hexbin}
\end{figure}

Since the average rate of resonances we have studied so far has effectively obliterated any differences in the corresponding temporal distribution, we can ask another important question: Are there any distinctive moments in the evolution of the simulated EEVs during which the systems experienced temporally higher resonance rates? After visually inspecting hundreds of resonance curves, we noticed that the aforementioned rate changes dramatically in many cases (cf.~the top panel of Fig.~\ref{fig:resonance-curves-sample} as an example).
In order to compare the temporal distribution of resonance events for the various EEVs we are dealing with, we performed this type of analysis on subgroups of systems divided according to the termination condition. We also normalised the time variable by dividing it by $T_{\rm max}$ of each resonance curve. This allowed us to present the whole evolution of components on a convenient and uniform interval, $[0,1]$. Figure~\ref{fig:time-series-of-resonances-hexbin} shows the results obtained for the primary components that have managed to deplete hydrogen in their cores. 

\begin{figure}
\centering
\includegraphics[width=1\hsize]{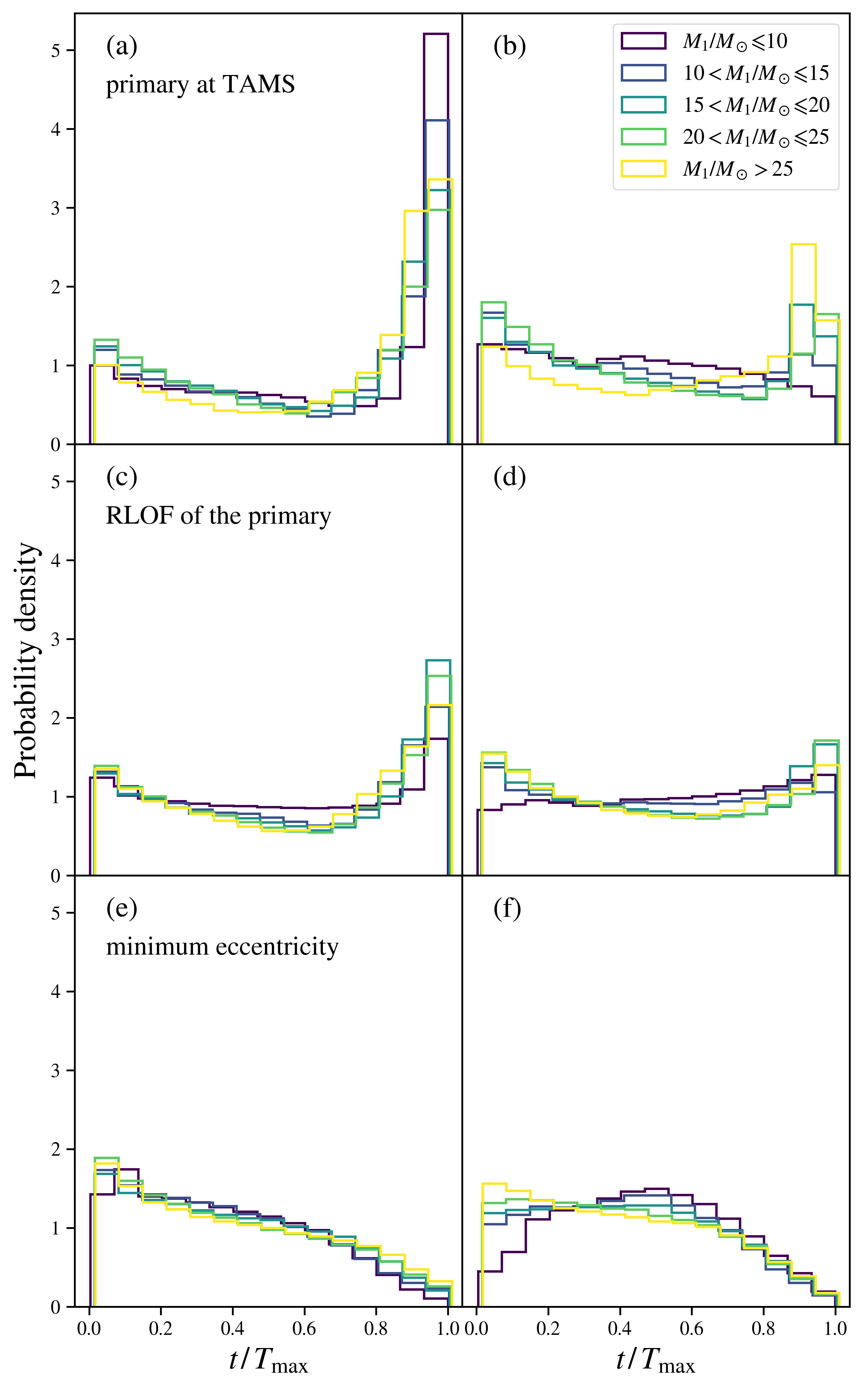}
\caption{Histograms of the normalised times of resonances occurring in the primary (left column) and secondary (right column) components. The consecutive rows (from top to bottom) correspond to EEVs satisfying different termination conditions, as labelled in panels (a), (c), and (e). The colour of the histogram is related to the initial mass range of the primary and is described in the legend in panel (b). We note that the histograms on the right (corresponding to the secondaries) refer to the different mass ranges of the primary component, not the secondary. For example, the yellowish histogram in panel (b) summarises the behaviour of all secondaries of the systems with the primaries having mass $M_1>25\,M_\odot$, i.e.~without distinguishing the mass ranges of $M_2$. The range of the ordinate axes is the same in each panel.}
\label{fig:histograms-of-resonances}
\end{figure}

Figure~\ref{fig:time-series-of-resonances-hexbin} also demonstrates that the distribution discussed here is not uniform over time. Specific areas in this diagram are clearly distinguishable. Nevertheless, this figure still contains information on the total number of resonances, which makes it somewhat problematic to compare the shapes of these distributions for different masses of the components. We have, therefore, prepared histograms of the times of resonances for five intervals of the primary's initial mass (every $5\,M_\odot$). Separate sets of histograms were generated for the primary and secondary components and the three main termination conditions\footnote{We did not prepare separate histograms for the EEVs, whose calculations were terminated due to the maximum allowed rotation rate. The size of this group (only eight systems) was insufficient for this task.}. All histograms are shown in Fig.~\ref{fig:histograms-of-resonances}.

The most diverse structure of the temporal distribution of resonances is shown by systems in which the primary component has completed its evolution in our simulations at TAMS (Fig.~\ref{fig:histograms-of-resonances}a and b). In fact, for all mass ranges, the distribution has two distinct maxima. The smaller of the two is located near the ZAMS, while the other is just before reaching the TAMS. Their presence can be explained by the rate of change in the stellar eigenspectrum, which is the highest (after averaging over all modes) at the aforementioned moments of evolution. In particular, the rapid changes in the radius of the star when it is close to complete depletion of hydrogen in its core cause a very high `concentration' of resonances in the final MS phase. The height of this dominant maximum decreases towards higher masses, but at the same time it becomes wider and wider. The distribution for the secondary components also reveals this kind of maximum near the TAMS, which is particularly well pronounced for companions of primaries with masses $\gtrsim25\,M_\odot$. Given these facts, an interesting conclusion can be drawn. Massive and intermediate-mass EEVs with at least one component leaving the MS should experience an increased rate of encountered resonances. One might therefore suspect that there is a statistically higher chance of observing TEOs in more evolved EEVs. The properties of the histograms for EEVs in which RLOF eventually occurred at the periastron (Fig.~\ref{fig:histograms-of-resonances}c and d) are very similar to the case described above. The only evident difference between the two is the reduction in maximum of the distribution near TAMS. This is due to the fact that the primary is likely to start the RLOF earlier than it reaches the TAMS, preventing the occurrence of a large number of resonances in a relatively short time, as mentioned earlier.

Eccentric systems that are subject to effective circularisation (Fig.~\ref{fig:histograms-of-resonances}e and f) behave quite differently from the two previous cases. They experience the vast majority of their resonance phenomena at the beginning of evolution, and then reduce the number of resonances almost monotonically, as the orbital eccentricity becomes smaller and smaller with time. Hence, the chance of observing TEOs in initially relatively tight EEVs (cf.~Fig.~\ref{fig:P-e-final-plane}a) is largest in the vicinity of ZAMS, which stays in contrast to the systems described above.

\subsection{Investigation of the morphology of resonance curves using UMAP}\label{sect:results-Investigation of the morphology of resonance curves using UMAP}
All the analysis described above was based solely on the distribution of resonances in time, i.e. neglecting the actual morphology of the resonance curves, e.g. differences in the height and width of resonance maxima, mean level of $\mathcal{L}(t)$, long-term trends in $\mathcal{L}(t)$, etc. Using the dimensionality reduction techniques presented in Sect.~\ref{sect:methods-ML analysis of the resonance curves}, we constructed 2D UMAP embeddings of the space of resonance curves in terms of their morphological features. Figures~\ref{fig:umap_l1_tot_ttmax} and \ref{fig:umap_l2_tot_ttmax} show the results obtained for the resonance curves of the primary and secondary component, respectively. We recall that the idea of the low-dimensional embedding performed here is to preserve the distances between two points in the original space as accurately as possible, so that the distances in the 2D plane reflect the distances in the full (original) space of morphological features (the vector of 2,000 quantiles, $\vec{Q}$). In other words, a pair of distant points in Figs.~\ref{fig:umap_l1_tot_ttmax} and \ref{fig:umap_l2_tot_ttmax} should correspond to resonance curves with notably different morphologies and ,\emph{vice versa}, a pair of resonance curves with similar properties is expected to lie in mutual vicinity on the 2D UMAP plane. Thanks to this key property of UMAP and many other dimensionality reduction methods we can effectively explore the entire space of resonance curve morphologies.

\subsubsection{UMAP plane for primary components}
\begin{figure*}
   \centering
   \includegraphics[width=0.95\hsize]{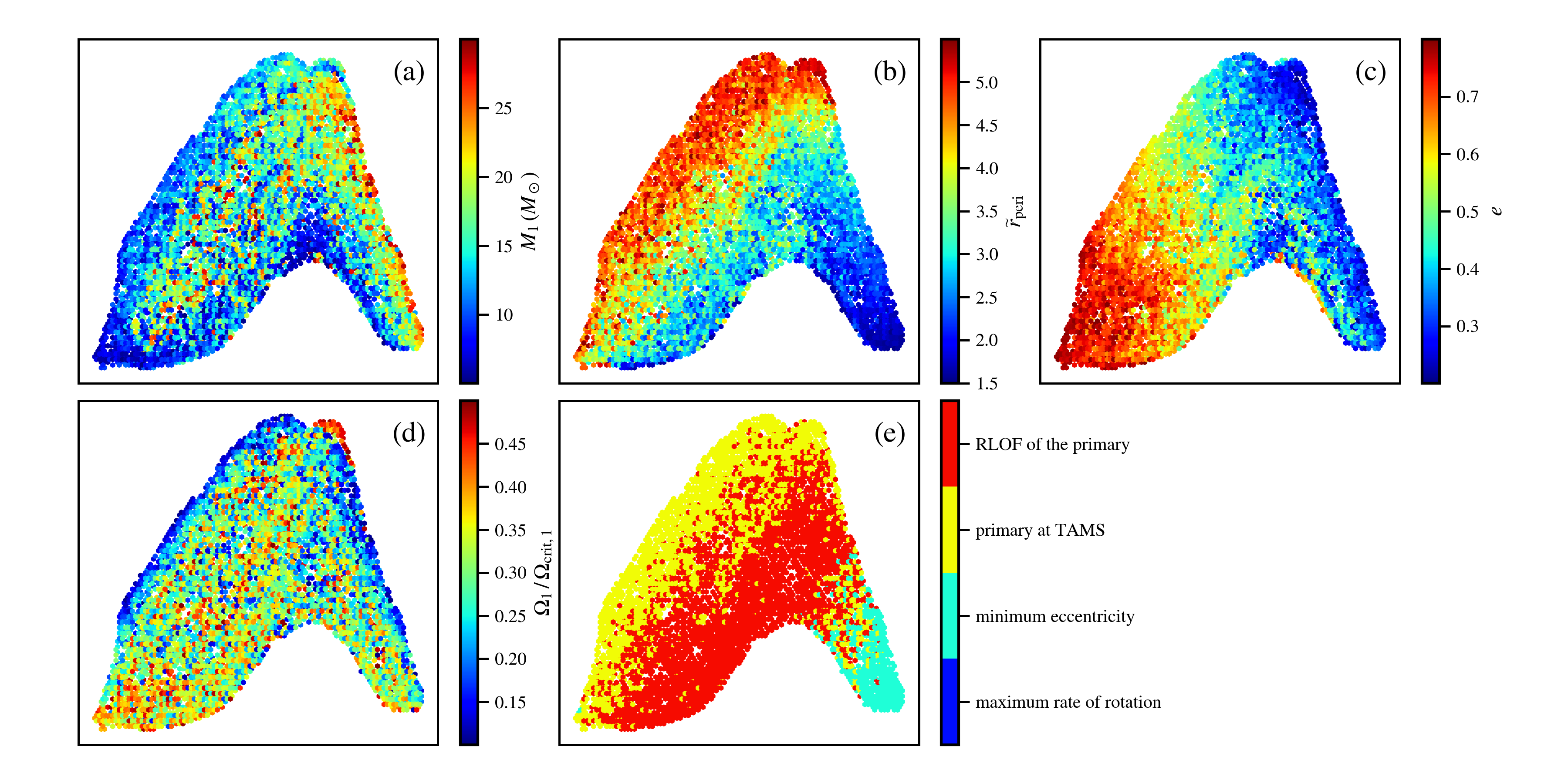}
   \caption{2D UMAP embedding of the manifold spanned by the morphological features of the resonance curves of the primary components. For details on how to obtain the presented embedding, see Sect.~\ref{sect:methods-ML analysis of the resonance curves}. Panels (a)\,--\,(d) are colour-coded with respect to the initial parameters of the simulated EEVs, as shown on the corresponding colour bars. The other initial parameters were omitted as they were not significantly related to the location of the points on the presented map. The different colours of points in panel (e) correspond to the termination condition, as shown in the legend on the right. The values on the abscissa and ordinate axes were omitted as they have no physical meaning. For clarity, the colour-coded features have been averaged within the small hexagonal areas in each panel. A discussion of the figure can be found in Sect.~\ref{sect:results-Investigation of the morphology of resonance curves using UMAP}.}
   \label{fig:umap_l1_tot_ttmax}
\end{figure*}

\begin{figure*}
   \centering
   \includegraphics[width=0.95\hsize]{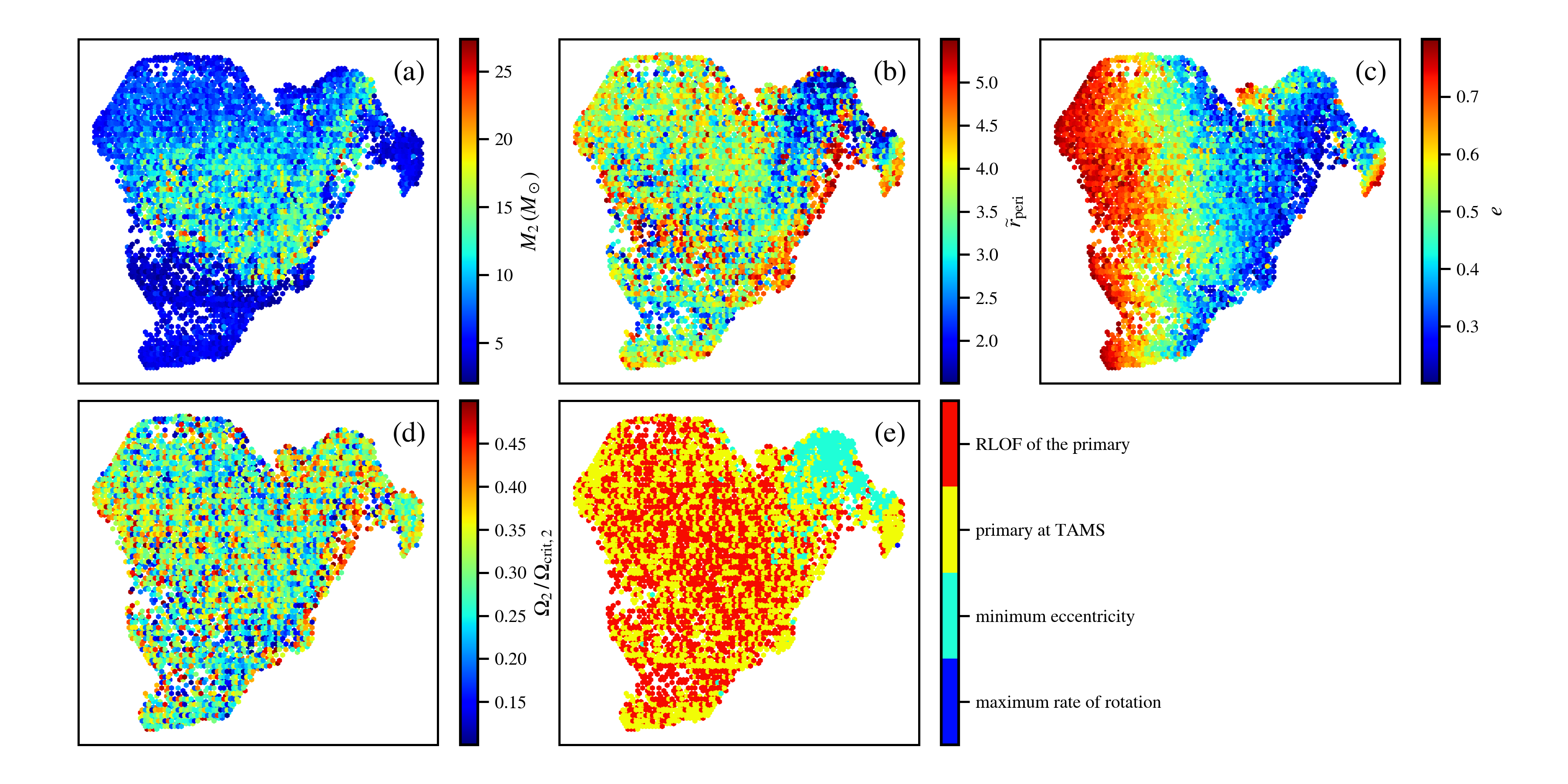}
   \caption{Same as Fig.~\ref{fig:umap_l1_tot_ttmax}, but for a set of resonance curves of the secondary components.}
   \label{fig:umap_l2_tot_ttmax}
\end{figure*}

We begin with a discussion of Fig.~\ref{fig:umap_l1_tot_ttmax}. Firstly, the presented 2D embedding does not indicate the presence of any well-separated groups among the resonance curves for the primary components. This is an observation that is true over the entire range of different values of the UMAP free parameters (Appendix~\ref{appendix:umap}) as well as for the different summary statistics of the resonance curves that were considered during the preliminary experiments. The morphology of the resonance curves changes smoothly depending on to the initial parameters of the simulated EEVs.

Secondly, as can be seen in Figs.~\ref{fig:umap_l1_tot_ttmax}b and c, the initial eccentricity and normalised periastron distance are parameters strongly correlated with the overall morphology of the resonance curves of the primary components. Moreover, their gradients in the UMAP plane are approximately orthogonal. Therefore, the pair of these parameters is the primary factor that determines the shape of $\mathcal{L}_1(t)$. The termination condition (Fig.~\ref{fig:umap_l1_tot_ttmax}e) generally follows the behaviour of $\widetilde{r}_{\rm peri}$ except at small periastron distances, when the morphology remains similar but the simulations were terminated due to hydrogen depletion in the primary's core or near-complete circularisation of the orbit. The initial mass of the primary component (Fig.~\ref{fig:umap_l1_tot_ttmax}a) and its initial angular velocity of rotation (Fig.~\ref{fig:umap_l1_tot_ttmax}d) are second-order factors shaping the morphology of the $\mathcal{L}_1(t)$ resonance curves. In the inner part of the plane, $M_1$ is distributed almost randomly. The clear exception is the boundary of the plane which can be roughly divided into two parts of mostly high or low initial mass of the primary. A similar conclusion can be drawn for the initial $\Omega_1/\Omega_{\rm crit,1}$. This time, however, the upper right part of Fig.~\ref{fig:umap_l1_tot_ttmax}d reveals a well-defined group of high initial $\Omega_1\,/\,\Omega_{\rm crit,1}$ and $\widetilde{r}_{\rm peri}$.

\begin{figure*}
   \centering
   \includegraphics[width=0.95\hsize]{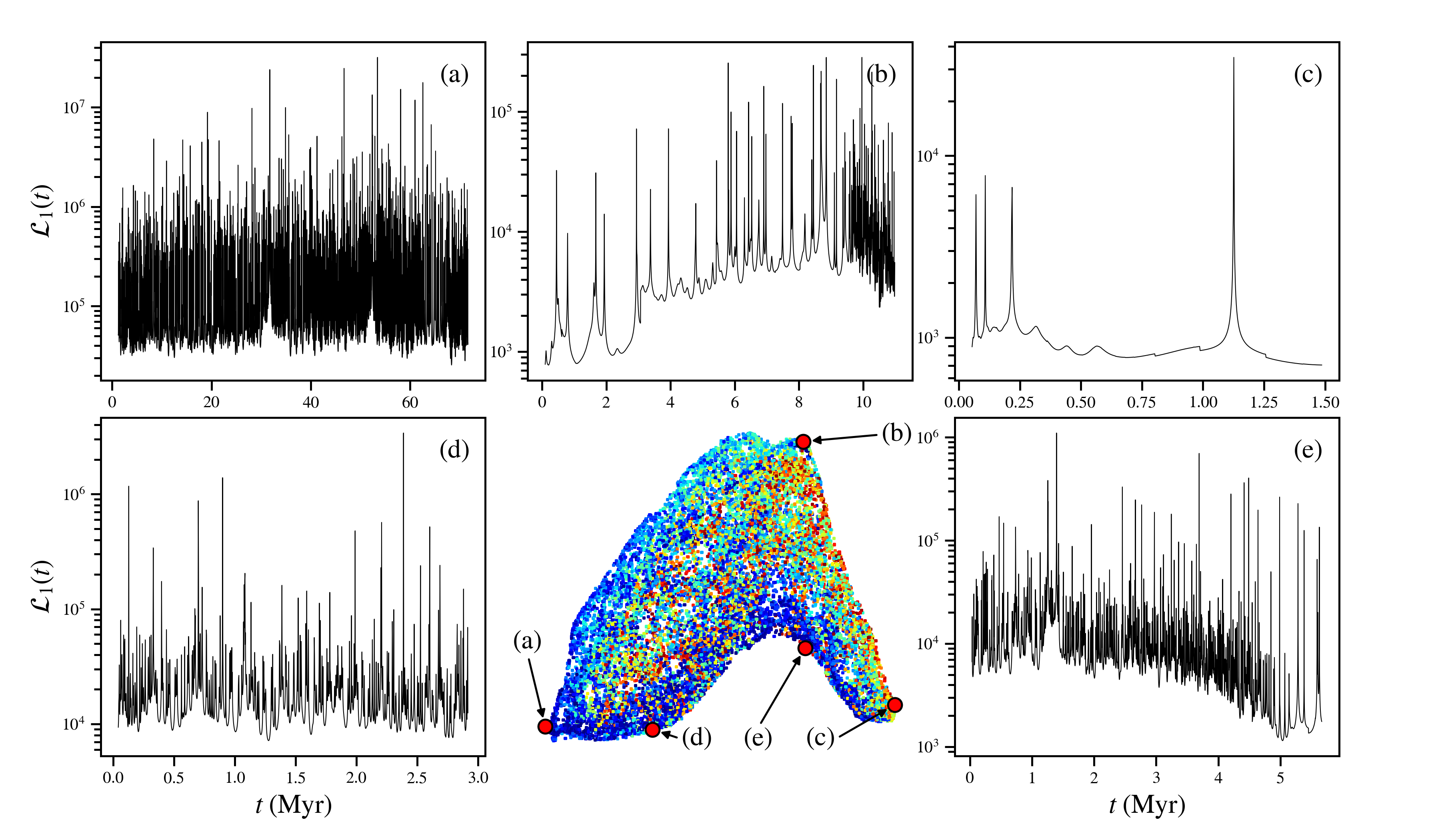}
   \caption{Variations in the morphology of the resonance curve for the primary component across the 2D UMAP plane from Fig.~\ref{fig:umap_l1_tot_ttmax}. The middle panel in the bottom row shows the plane with colour-coding identical to that in Fig.~\ref{fig:umap_l1_tot_ttmax}a (without hexagonal binning). Panels (a)\,--\,(e), which surround the area, show example resonance curves that correspond to the locations on the area masked with large red dots and labelled according to the associated panel. The positions of points (a)\,--\,(d) have been chosen in such a way as to correspond to different extreme positions in the plain, while point (e) refers to one of the intermediate cases. A discussion of the figure can be found in Sect.~\ref{sect:results-Investigation of the morphology of resonance curves using UMAP}.}
   \label{fig:umap_with_examples_l1_with_examples_ttmax}
\end{figure*}

It is difficult to include here a complete presentation of the changes in the morphology of $\mathcal{L}_1(t)$ as a function of their position on the UMAP plane. Therefore, we only focus on some extreme points to present some boundary cases. Figure~\ref{fig:umap_with_examples_l1_with_examples_ttmax} shows examples of $\mathcal{L}_1(t)$ from different areas of the morphological plane. Primary components with lower masses and high initial eccentricities are generally characterised by resonance curves with high mean levels and a rich set of resonances, as shown in Fig.~\ref{fig:umap_with_examples_l1_with_examples_ttmax}a. The resonance curve depicted in Fig.~\ref{fig:umap_with_examples_l1_with_examples_ttmax}b represents intermediate-mass fast-rotating primary with large initial $\widetilde{r}_{\rm peri}$ and low initial eccentricity. Here, the base level of $\mathcal{L}_1(t)$ increases by an order of magnitude and then the system experiences a large number of resonances, during the evolution near TAMS. The increase in the mean value of $\mathcal{L}_1(t)$ is characteristic of stars with a high initial rotation rates. Primary components lying between points (a) and (b) generally do not manifest this characteristic. Moving along a straight line on the plane from (a) to (b), the increase in the number of resonances during the evolution near TAMS becomes more and more apparent. Figure~\ref{fig:umap_with_examples_l1_with_examples_ttmax}c shows an example of a system with a small initial eccentricity and a short periastron distance at ZAMS that is rapidly circularising. As expected, the $\mathcal{L}_1(t)$ resonance curves for such objects have a small number of resonance maxima and a low base level. The case corresponding to the larger initial eccentricity is shown in Fig.~\ref{fig:umap_with_examples_l1_with_examples_ttmax}e, where the total number of resonances is much greater. In this case, the evolution is mainly distinguished by a decrease in the frequency of resonances with time, related to the efficient circularisation of the orbit, and therefore a decrease in the mean level of $\mathcal{L}_1(t)$. Finally, the resonance curve in Fig.~\ref{fig:umap_with_examples_l1_with_examples_ttmax}d is representative of the most EEVs between points (c) and (d). They are characterised by an approximately uniform distribution of resonances over time and an almost constant base level of the resonance curve.

\begin{figure*}
   \centering
   \includegraphics[width=0.95\hsize]{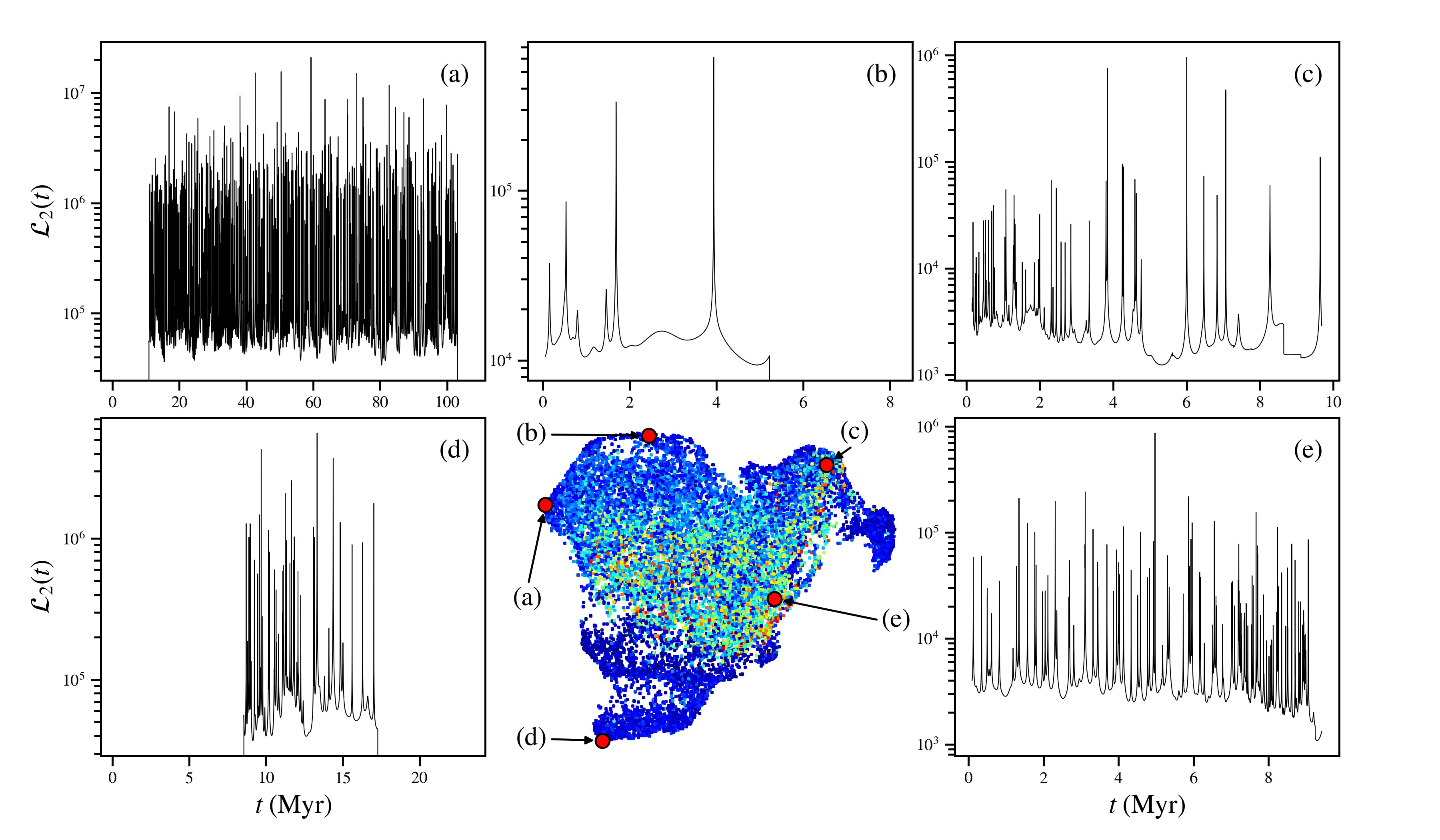}
   \caption{Same as Fig.~\ref{fig:umap_with_examples_l1_with_examples_ttmax}, but for a set of resonance curves of the secondary components.}
   \label{fig:umap_with_examples_l2_with_examples_ttmax}
\end{figure*}

\subsubsection{UMAP plane for the secondary components}
The situation for the secondary component (Fig.~\ref{fig:umap_l2_tot_ttmax}) is quite different from the previous case. The UMAP manifold obtained for the set of $\mathcal{L}_2(t)$ reveals slightly more complex structure than the shape of embedding in Fig.~\ref{fig:umap_l1_tot_ttmax}. Since the time span of $\mathcal{L}_2(t)$ is largely determined by the mass of the primary component, the resonance curves for the secondary components were terminated at times not necessarily related to their actual evolutionary status and are statistically shorter than they could be for primaries of the same mass. For this reason, secondary components experience, on average, fewer resonances, but, at the same time, their resonance curves can take more diverse forms compared to $\mathcal{L}_1(t)$. Undoubtedly, the main factor shaping the morphology of the $\mathcal{L}_2(t)$ is the initial eccentricity (Fig.~\ref{fig:umap_l2_tot_ttmax}c), which with the exception for two small areas, varies smoothly across the UMAP plane. The other parameters (Fig.~\ref{fig:umap_l2_tot_ttmax}a, b and d) play a secondary role, showing the complex and fine structures on the plane. As can easily be seen in Fig.~\ref{fig:umap_l2_tot_ttmax}a, the extreme cases of $\mathcal{L}_2(t)$ in terms of their morphology belong almost exclusively to the EEVs with intermediate-mass secondary components hat gather at the periphery of the plane. Some of these objects even form slightly better separated groups, isolating from the central part of the area.

Five limiting examples of resonance curves for secondary components are shown in Fig.~\ref{fig:umap_with_examples_l2_with_examples_ttmax}. Panels (a), (c) and (e) together with the area they approximately enclose contain resonance curves morphology very similar to that described for primary components. The resonance curves belonging to the `clouds' of points labelled as (b) and (d) in Fig.~\ref{fig:umap_with_examples_l2_with_examples_ttmax} are completely different. They are distinguished by the complete absence of resonances during a certain period of the evolution of the system. The differentiating feature of these cases is the disappearance of resonances from some time to the end of evolution (Fig.~\ref{fig:umap_with_examples_l2_with_examples_ttmax}b) or the presence of resonances only around the middle of the considered evolution time (Fig.~\ref{fig:umap_with_examples_l2_with_examples_ttmax}d).

%-------------------------------------------------------------------
\section{Summary and conclusions}\label{sect:summary-and-conclusions}
%Massive stars play a key role in our understanding of the Universe as they are progenitors of many energetic and exotic phenomena, for instance, their outcomes include several types of supernovae and gravitational wave events. As widely known, massive and intermediate-mass MS stars often reside in the binary or even multiple systems. Additionally, due to their relatively young age and radiative envelopes, the systems which they constitute are commonly characterised by significantly eccentric orbits. Thus, many of the binary systems that contain massive or intermediate-mass component(s) can be observed as EEVs (Sect.~\ref{sect:introduction}). In parallel to the equilibrium tide responsible for the `heartbeat' in the light curve of an EEV, the orbital phase-dependent tidal potential acting on the components of EEV can induce TEOs in their interiors, which in turn can affect the evolution of the binary system (Sect.~\ref{sect:properties of TEOs}). Most TEOs are damped normal modes, meaning that without constant tidal forcing they would not be observed in the star. This property makes them a valuable tool for seismic studies of the massive and intermediate-mass stars without any naturally overstable modes, allowing to perform so-called tidal asteroseismology for them.

In our paper, we aimed to investigate the temporal variation of conditions that favour excitation of TEOs in EEVs with massive and intermediate-mass MS components (Sect.~\ref{sect:properties of TEOs}) and see how their picture changes with different initial parameters of the system. In order to achieve this goal, we simulated the evolution of 20,000 EEVs using the \texttt{MESA} software in combination with the \texttt{GYRE} stellar oscillations code (Sect.~\ref{sect:methods}). Our calculations started at ZAMS and were terminated if one of the conditions presented in Sect.~\ref{sect:methods-termination} was met. We considered only modes with $l=2$, $m=0,+2$ because they are expected to be dominant TEOs. We also assumed that all TEOs are due to chance resonances, i.e. we neglected the effect of TEO on the orbit. Knowing the evolution of the orbital parameters of simulated EEVs and the temporal changes in the eigenmode spectra of the components, we were able to derive resonance curves $\mathcal{L}_1(t)$ and $\mathcal{L}_2(t)$ defined by Eqs.~(\ref{eqn:L(t)}) and (\ref{eq:L_N}). The equations reflect the overall resonance conditions, and thus indirectly also the chance of TEOs, separately for the primary and secondary components of our simulated EEVs.

After visually inspecting the obtained resonance curves, calculating basic statistics for them and applying ML-based methods to the entire data set, our main results can be summarised as follows.
\begin{enumerate}
    \item Resonance curves are characterised by striking diversity in terms of their morphology (Sect.~\ref{sect:results-eye-inspection}). EEV components can experience a very different number of resonances, and their distribution over time can take various forms, including the lack of resonances over a long periods of time. We also distinguished a group of resonance curves that exhibit prolonged resonances, about two orders of magnitude longer than typical (Sect.~\ref{sect:results-long duration resonances}, Fig.~\ref{fig:long-resonances}). These long resonances are the potential sources of high-amplitude and resonantly-locked TEOs.
    \item Resonances between tidal forcing frequencies and the spectrum of stellar normal modes are not rare events among massive and intermediate-mass MS EEVs (Sect.~\ref{sect:results-total-number-of-resonances}). Although the total number of resonances depends mostly on the initial orbital parameters, it is typically of the order of $10^2$\,--\,$10^3$ for a given system during the entire MS phase (Fig.~\ref{fig:total-number-of-resonances}). Let us emphasise at this point that these numbers are rather lower limits for the actual $\mathcal{N}_{\rm res}$ in EEVs because we considered only $l=2$ TEOs. Taking higher degree modes into account will certainly increase the reported values of $\mathcal{N}_{\rm res}$.
    \item On average, the more massive a star is, the higher the rate of resonances it experiences (Sect.~\ref{sect:results-total-number-of-resonances}). For the most massive stars in our sample ($\approx30\,M_\odot$), the average rate of resonances can reach $\sim$\,$10^2\,$Myr$^{-1}$, which is approximately an order of magnitude higher than for intermediate-mass stars (Fig.~\ref{fig:number-of-resonances-per-myr}).
    \item The distribution of resonances over time is not homogeneous and depends  primarily on whether the system circularises before the primary reaches the TAMS or RLOF occurs at the periastron (Sect.~\ref{sect:results-Distribution of resonances over time}, Fig.~\ref{fig:histograms-of-resonances}). We noticed a particular moment in the evolution of our EEVs near the TAMS, when the components undergo an increased number of resonances in a relatively short time (Fig.~\ref{fig:histograms-of-resonances}a and b).
    \item The low-dimensional representation of the morphology of the resonance curves, summarised by quantile-based statistics and subsequently processed by UMAP, shows that its manifold forms a rather smooth distribution without well defined (separated) groups (Sect.~\ref{sect:results-Investigation of the morphology of resonance curves using UMAP}, Figs.~\ref{fig:umap_l1_tot_ttmax} and \ref{fig:umap_l2_tot_ttmax}). Less differentiated, at least in terms of the adopted method, are the resonance curves of the primary components, for which the initial eccentricity and the normalised periastron distance largely determine their morphological features. Although secondary components experience far fewer resonances, their shapes are generally more complex due to the predominant influence of the primary component on evolution time.
\end{enumerate}

In light of the results obtained in our study, we can draw several interesting conclusions. Firstly, statistically speaking, TEOs are more likely to be discovered in more massive EEVs, as their components have a higher average rate of resonances. This does not necessarily mean that a higher absolute number of more massive EEVs exhibiting TEOs than less massive ones will be observed\footnote{Due to the rapid decrease of the mass function towards larger stellar masses \citep[e.g.,][]{2003PASP..115..763C}.}. However, when comparing two particular systems, one with intermediate-mass components and the other with much higher masses of the components, it is for the latter that we have a statistically higher chance that some resonance is currently underway there. Secondly, it seems that TEOs should be especially well visible in EEVs that contain a component approaching TAMS. Given these facts, the `extreme-amplitude' massive EEV, MACHO\,80.7443.1718 \citep{2021MNRAS.506.4083J,2022A&A...659A..47K} fits this picture almost perfectly. Its primary component is a B0.5\,Ib-II supergiant leaving the MS and, more importantly, many high-amplitude TEOs have now been detected in this extreme system. It is possible that what the primary component of MACHO\,80.7443.1718 is currently undergoing corresponds to the resonance curve shown in Fig.~\ref{fig:umap_with_examples_l1_with_examples_ttmax}b, i.e. it is in the phase of a high resonance rate caused by relatively fast changes in its radius and orbital parameters. Moreover, the amplitudes of TEOs observed in MACHO\,80.7443.1718 vary over time notably, suggesting that we may witness rapid changes in the resonance conditions for the primary component of this particular EEV. It would therefore be very valuable to carry out an observational study for a large sample of EEVs to verify whether TEOs are common in massive and intermediate-mass EEVs whose components have already depleted most of the hydrogen in their cores. With high-quality space-borne photometric observations, both operational, such as the Transiting Exoplanet Survey Satellite \citep{2015JATIS...1a4003R} or BRITE-Constellation \citep{2014PASP..126..573W}, and planned missions \citep[e.g. Planetary Transits and Oscillations of Stars,][]{2014ExA....38..249R}, it is definitely a feasible task.

The excitation of $g$-mode TEOs, which propagate deep inside the star, may be an underestimated mechanism for angular momentum (AM) transport inside the components of EEVs. It has long been suspected that self-excited oscillations and internal gravity waves\footnote{Gravity waves which are stochastically driven by the turbulent convective motions near the interface of the convective core and the envelope \citep[see][for a recent review]{2020A&A...640A..36B}.} efficiently redistribute AM in the radial direction of the star \citep[e.g.,][]{2013ApJ...772...21R,2017ApJ...848L...1R}. Although the majority of our resonances have relatively short durations (of the order of $10^3$\,--\,$10^4$\,years), they can be quite frequent (especially near the TAMS), hence the question of their contribution to AM transport and mixing processes becomes urgent for the components of EEVs. Performing calculations that would treat the evolution of the orbit, components and TEOs in a fully self-consistent way seems particularly interesting for massive eccentric systems leaving the MS.

We have already entered the era of observational studies of distant star-bursting galaxies and stellar populations in low-metallicity environments, that shaped the Universe in its early epochs. Recalling that the metal-poor stars were much more massive than their current metal-rich counterparts \citep[e.g.,][]{Hosokawa_2013,2014ApJ...792...32S}, we can ask what effect metallicity has on the occurrence of TEOs in massive EEVs and how the results we presented depend on metals content. Therefore, future studies of the importance of TEOs in massive, metal-poor EEVs seems worthy further investigation, especially because of the ongoing James Webb Space Telescope mission\footnote{Among numerous possibilities of JWST is also the possibility of resolving the most massive (and luminous) EEVs in the Local Group of galaxies, including nearby very metal-poor dwarf galaxies \citep[e.g., Sextans~A dwarf galaxy,][the metallicity of which amounts to about $1/10\,$Z$_\odot$]{Kaufer_2004}.} \citep[JWST,][]{2006SSRv..123..485G}, which is certain to bring many discoveries in the stellar astrophysics of early stellar populations, including stars in eccentric binary systems.

%-------------------------------------------------------------------

\begin{acknowledgements}
PKS is indebted to his brother, Adam Karol Ko{\l}aczek-Szyma\'nski, who oversaw the purchase and assembly of a dedicated PC workstation to enable the efficient calculation of models in \texttt{MESA} and \texttt{GYRE}. Without his generous help, this project would literally never have been completed.\\

The authors are thankful to Prof.~Andrzej Pigulski for many important suggestions and fruitful discussions that made this manuscript more comprehensible, and to the anonymous referee for many inspiring comments that helped to improve the manuscript.\\

PKS was supported by the Polish National Science Center grant no.~2019/35/N/ST9/03805. TR was partly founded from budgetary funds for science in 2018-2022 in a research project under the program ,,Diamentowy Grant'', no. DI2018 024648. Much of this work was developed and written during IAU Symposium\,361, ,,Massive Stars Near \& Far'', held in Ballyconnell, Ireland, 8\,--\,13 May, 2022. PKS is very grateful to the organisers for the opportunity to participate in this event.

\end{acknowledgements}


\bibliographystyle{aa}
\bibliography{bib}

\begin{appendix}

\section{\texttt{MESA} input files}\label{appendix:mesa}
\texttt{MESAbinary} needs at least three input files (hereafter `inlists') to start the calculations. Two of them provide all the necessary parameters to perform the evolution of each component separately\footnote{Details of each keyword in \texttt{MESAstar} v.\,r15140 inlist can be found at \url{https://docs.mesastar.org/en/r15140/reference/star_job.html} and \url{https://docs.mesastar.org/en/r15140/reference/controls.html}}, and the last one describes the evolution of the orbit as well as other processes that depend on binarity\footnote{Details of each keyword in \texttt{MESAbinary} v.\,r15140 inlist can be found at \url{https://docs.mesastar.org/en/r15140/reference/binary_job.html} and \url{https://docs.mesastar.org/en/r15140/reference/binary_controls.html}} (e.g.~spin-orbit coupling). In the following appendix, we present example \texttt{MESAstar} and \texttt{MESAbinary} inlists, which we used to generate a set of the evolutionary tracks of the components of the binary system. However, we have intentionally omitted any controls related to the names of the files or directories where the results should be stored. All values that changed in the inlists depending on the simulated binary system were enclosed in square brackets -- [\,$\cdot$\,]. The parameters adopted below resulted in a typical number of about 2,400 zones in the radial direction of the star. The number of models calculated per single EEV was typically of the order of several hundred, mainly depending on the termination condition.

\subsection{\texttt{MESAstar} inlist}
{\tt
   \&star\_job\\
   !OUTPUT\\
   history\_columns\_file = "my\_history\_columns.list"\\
   profile\_columns\_file = "my\_profile\_columns.list"\\
   show\_log\_description\_at\_start = .false.\\
   save\_photo\_when\_terminate=.false.\\
   !MODIFICATIONS TO MODEL\\
   new\_rotation\_flag=.true.\\
   change\_rotation\_flag=.true.\\
   new\_omega\_div\_omega\_crit=[$\Omega/\Omega_{\rm crit}$]\\
   num\_steps\_to\_relax\_rotation=100\\
   relax\_omega\_max\_yrs\_dt = 1d4\\
   relax\_omega\_div\_omega\_crit=.true.\\
   set\_initial\_cumulative\_energy\_error = .true.\\
   new\_cumulative\_energy\_error = 0d0\\
   / ! end of star\_job namelist\\
   \&eos\\
   / ! end of eos namelist\\
   \&kap\\
   use\_Type2\_opacities = .true.\\
   Zbase = 0.02\\
   / ! end of kap namelist\\
   \&controls\\
   !SPECIFICATIONS FOR STARTING MODEL\\
   initial\_z=0.02d0\\
   !CONTROLS FOR OUTPUT\\
   terminal\_interval=100\\
   write\_header\_frequency=1\\
   photo\_interval=100000\\
   history\_interval=5\\
   star\_history\_dbl\_format = "(1pes40.6e3, 1x)"\\
   profile\_interval=10\\
   max\_num\_profile\_models=5000\\
   write\_pulse\_data\_with\_profile=.true.\\
   pulse\_data\_format="GYRE"\\
   add\_double\_points\_to\_pulse\_data=.true.\\
   !WHEN TO STOP\\
   max\_model\_number = 5000\\
   xa\_central\_lower\_limit\_species(1)="h1"\\
   xa\_central\_lower\_limit(1)=1d-4\\
   omega\_div\_omega\_crit\_limit=0.75\\
   !MIXING PARAMETERS\\
   mixing\_length\_alpha=1.82d0\\
   use\_Ledoux\_criterion=.true.\\
   num\_cells\_for\_smooth\_gradL\_composition\_term = 0\\
   alpha\_semiconvection=0.01d0\\
   okay\_to\_reduce\_gradT\_excess=.true.\\
   mlt\_make\_surface\_no\_mixing = .true.\\
   overshoot\_scheme(1)="exponential"\\
   overshoot\_zone\_type(1) = "burn\_H"\\
   overshoot\_zone\_loc(1) = "core"\\
   overshoot\_bdy\_loc(1) = "top"\\
   overshoot\_f(1) = [$f_{\rm ov}$]\\
   overshoot\_f0(1) = 0.005\\
   do\_conv\_premix=.true.\\
   set\_min\_D\_mix=.true.\\
   min\_D\_mix=1d5\\
   !ROTATION CONTROLS\\
   am\_D\_mix\_factor=0.0333333d0\\
   D\_DSI\_factor = 1\\
   D\_SH\_factor = 1\\
   D\_SSI\_factor = 1\\
   D\_ES\_factor = 1\\
   D\_GSF\_factor = 1\\
   !ATMOSPHERE BOUNDARY CONDITION\\
   atm\_option="table"\\
   atm\_table="photosphere"\\
   !MASS GAIN OR LOSS\\
   hot\_wind\_scheme="Vink"\\
   hot\_wind\_full\_on\_T=1.2d4\\
   cool\_wind\_full\_on\_T=0.9d3\\
   Vink\_scaling\_factor=1d0\\
   no\_wind\_if\_no\_rotation=.true.\\
   mdot\_omega\_power=0.43d0\\
   max\_mdot\_jump\_for\_rotation=5d0\\
   rotational\_mdot\_kh\_fac = 1.0d3\\
   !MESH ADJUSTMENT\\
   max\_delta\_x\_for\_merge = 0.01d0\\
   max\_dq=1d-3\\
   min\_dq=1d-16\\
   min\_dq\_for\_split=1d-16\\
   !ASTEROSEISMOLOGY CONTROLS\\
   num\_cells\_for\_smooth\_brunt\_B = 0\\
   !STRUCTURE EQUATIONS\\
   use\_dedt\_form\_of\_energy\_eqn = .true.\\
   !TIMESTEP CONTROLS\\
   min\_timestep\_factor=0.5d0\\
   max\_timestep\_factor=2.0d0\\
   dH\_div\_H\_limit=0.5d0\\
   delta\_lgL\_phot\_limit = 0.05d0\\
   / ! end of controls namelist
}

\subsection{\texttt{MESAbinary} inlist}
{\tt
   \&binary\_job\\
   !OUTPUT/INPUT FILES\\
   show\_binary\_log\_description\_at\_start = .false.\\
   binary\_history\_columns\_file = "my\_binary\_history\_columns.list"\\
   !STARTING MODEL\\
   evolve\_both\_stars=.true.\\
   change\_ignore\_rlof\_flag = .true.\\
   new\_ignore\_rlof\_flag = .true.\\
   / ! end of binary\_job namelist\\
   \&binary\_controls\\
   !SPECIFICATIONS FOR STARTING MODEL\\
   m1=[$M_1$]\\
   m2=[$M_2$]\\
   initial\_eccentricity=[$e$]\\
   initial\_period\_in\_days=-1\\
   initial\_separation\_in\_Rsuns=[$a$]\\
   !CONTROLS FOR OUTPUT\\
   history\_interval=5\\
   photo\_interval=100000\\
   terminal\_interval=100\\
   write\_header\_frequency=1\\
   !TIMESTEP CONTROLS\\
   fa=0.02d0\\
   fa\_hard=0.03d0\\
   fr=0.10d0\\
   fj=0.001d0\\
   fj\_hard=0.005d0\\
   fe=0.02d0\\
   fr\_dt\_limit = 1.0d2\\
   fdm = 1d-3\\
   fdm\_hard = 5d-3\\
   dt\_softening\_factor = 0.3d0\\
   varcontrol\_ms=5d-4\\
   varcontrol\_post\_ms=5d-4\\
   dt\_reduction\_factor\_for\_j=5d-2\\
   !MASS TRANSFER CONTROLS\\
   do\_enhance\_wind\_1=.true.\\
   do\_enhance\_wind\_2=.true.\\
   tout\_B\_wind\_1 = [$B_{\rm wind}$]\\
   tout\_B\_wind\_2 = [$B_{\rm wind}$]\\
   !ORBITAL JDOT CONTROLS\\
   do\_jdot\_gr=.true.\\
   do\_jdot\_ls=.true.\\
   do\_jdot\_ml=.true.\\
   do\_jdot\_mb=.false.\\
   !ROTATION AND SYNC CONTROLS\\
   do\_tidal\_sync=.true.\\
   sync\_type\_1="Hut\_rad"\\
   sync\_type\_2="Hut\_rad"\\
   !ECCENTRICITY CONTROLS\\
   do\_tidal\_circ=.true.\\
   circ\_type\_1="Hut\_rad"\\
   circ\_type\_2="Hut\_rad"\\
   anomaly\_steps=300\\
   / ! end of binary\_controls namelist
}

\section{\texttt{GYRE} input file}\label{appendix:gyre}
The \texttt{GYRE} stellar oscillations code requires a single input file that collects all the user-specified parameters of the asteroseismic calculations being performed\footnote{Details of each keyword in \texttt{GYRE} v.\,6.0.1 input file can be found at \url{https://gyre.readthedocs.io/en/v6.0.1/ref-guide/input-files.html}}. Below is our example file {\tt gyre.in}. As in Appendix~\ref{appendix:mesa}, we have omitted any keywords related to specific file names and have highlighted variables by enclosing them in square brackets.\\

   {\tt
   \noindent \&constants\\
   /\\
   \&model\\
   model\_type  = "EVOL"\\
   file\_format = "MESA"\\
   /\\
   \&mode\\
   l = 2\\
   m = 0\\
   n\_pg\_min = -30\\
   n\_pg\_max = 30\\
   tag      = "m0"\\
   /\\
   \&mode\\
   l = 2\\
   m = 2\\
   n\_pg\_min = -30\\
   n\_pg\_max = 30\\
   tag      = "m2"\\
   /\\
   \&osc\\
   inner\_bound  = "REGULAR"\\
   outer\_bound  = "VACUUM"\\
   adiabatic    = .true.'\\
   nonadiabatic = .true.'\\
   /\\
   \&rot\\
   coriolis\_method = "TAR"\\
   Omega\_rot\_source = "MODEL"\\
   /\\
   \&num\\
   ad\_search = "BRACKET"\\
   nad\_search = "AD"\\
   diff\_scheme = "MAGNUS\_GL2"\\
   /\\
   \&scan\\
   grid\_type  = "LINEAR"\\
   freq\_min   = [$f_{\rm min}^{m=0}$]\\
   freq\_max   = [$f_{\rm max}^{m=0}$]\\
   n\_freq     = [$N_{\rm freq}^{m=0}$]\\
   freq\_units = "CYC\_PER\_DAY"\\
   grid\_frame = "INERTIAL"\\
   freq\_frame = "INERTIAL"\\
   tag\_list       = "m0"\\
   /\\
   \&scan\\
   grid\_type  = "LINEAR"\\
   freq\_min   = [$f_{\rm min}^{m=+2}$]\\
   freq\_max   = [$f_{\rm max}^{m=+2}$]\\
   n\_freq     = [$N_{\rm freq}^{m=+2}$]\\
   freq\_units = "CYC\_PER\_DAY"\\
   grid\_frame = "COROT\_I"\\
   freq\_frame = "COROT\_I"\\
   tag\_list   = "m2"\\
   /\\
   \&grid\\
   /\\
   \&ad\_output\\
   /\\
   \&nad\_output\\
   summary\_file\_format = "TXT"\\
   summary\_item\_list   = "freq,l,m,n\_p,n\_g,n\_pg"\\
   freq\_units          = "CYC\_PER\_DAY"\\
   freq\_frame          = "INERTIAL"}\\
   
The frequency scan limits, $f_{\rm min}^{m=0}$, $f_{\rm max}^{m=0}$, $f_{\rm min}^{m=+2}$, and $f_{\rm max}^{m=+2}$, were calculated as described in Sect.~\ref{sect:asteroseismic-calculations}. The total numbers of discrete frequency points, $N_{\rm freq}^{m=0}$, and $N_{\rm freq}^{m=+2}$, were obtained as follows,
\begin{equation}
    N_{\rm freq}^{m=0,+2} = \lceil (f_{\rm max}^{m=0,+2}-f_{\rm min}^{m=0,+2})/(0.005\,{\rm d}^{-1}) \rceil,
\end{equation}
where $\lceil\,\cdot\,\rceil$ denotes the ceiling function.

\section{Data used by \texttt{MESA}}\label{appendix:mesa-data}
Our work uses the \texttt{MESA} stellar evolution code, which incorporates a vast compilation of knowledge, mainly from micro- and macrophysics, collected by many authors. The \texttt{MESAeos} module is a mixture of OPAL \citep{Rogers2002}, SCVH \citep{Saumon1995}, FreeEOS \citep{Irwin2004}, HELM \citep{Timmes2000}, PC \citep{Potekhin2010} and Skye \citep{Jermyn2021} equation of states. Radiative opacities are taken primarily from OPAL \citep{Iglesias1993, Iglesias1996}, with low-temperature data from \citet{Ferguson2005} and the high-temperature, Compton-scattering dominated regime by \citet{Poutanen2017}.  Electron conduction opacities are from \citet{Cassisi2007} and \citet{Blouin2020}. Nuclear reaction rates are from JINA REACLIB \citep{Cyburt2010}, NACRE \citep{Angulo1999} and additional tabulated weak reaction rates from \citet{Fuller1985}, \cite{Oda1994} and \cite{Langanke2000}. Screening is included via the prescription of \citet{Chugunov2007}. Thermal neutrino loss rates are taken from \citet{Itoh1996}. Roche lobe radii in binary systems are computed using the fit of \citet{Eggleton1983}.

\section{Adjustable parameters of the UMAP}\label{appendix:umap}
UMAP, as a highly flexible method, is prone to returning misleading results in the case of inappropriately set free parameters. On the one hand, they can lead to the appearance of spurious groups and, on the other hand, to the loss of finer topological structure. The vital UMAP parameters that need adjustment are \texttt{n\textunderscore neighbors}, \texttt{min\textunderscore dist}, \texttt{n\textunderscore components} and \texttt{metric}. The \texttt{n\textunderscore neighbors} parameter is the most important, as it controls the balance between the local and global structure present in the data that will be mapped to the embedding. We experimented with different values of this parameter ranging from 5 to 1,000 (the default is 15) and concluded that the resonance curves (summarised by the proposed statistics) always form a single group, almost independently of the choice of \texttt{n\textunderscore neighbors}. Later, \texttt{min\textunderscore dist} sets the minimum distance between two different points on the embedding. We tested its values from 0.0 to 0.5 and do not observe any significant effect on manifold. We took the last two parameters, namely \texttt{n\textunderscore components} that specifies the number of dimensions of the embedding, and \texttt{metric} specifying the metric used for similarity calculation, as default values. Finally, we used the following set of free parameters: $\texttt{n\textunderscore neighbors}=500$, $\texttt{min\textunderscore dist}=0.1$, $\texttt{n\textunderscore components}=2$ and $\texttt{metric} = \texttt{'euclidean'}$.

\end{appendix}

\end{document}